\documentclass[aps,nofootinbib,twocolumn,showpacs]{revtex4}
\usepackage{amsmath,amsfonts,amssymb}
\usepackage{enumitem}
\usepackage{graphicx}
\usepackage[caption=false]{subfig}

\begin{document}

\title{Tunable compression of template banks for fast gravitational-wave detection and localisation}
\author{Alvin J. K. Chua}
\email{ajkc3@ast.cam.ac.uk}
\affiliation{Institute of Astronomy, University of Cambridge, Madingley Road, Cambridge CB3 0HA, United Kingdom}
\author{Jonathan R. Gair}
\email{j.gair@ed.ac.uk}
\affiliation{School of Mathematics, University of Edinburgh, King's Buildings, Edinburgh EH9 3JZ, United Kingdom}
\date{\today}

\begin{abstract}
One strategy for reducing the online computational cost of matched-filter searches for gravitational waves is to introduce a compressed basis for the waveform template bank in a grid-based search. In this paper, we propose and investigate several tunable compression schemes for a general template bank. Through offline compression, such schemes are shown to yield faster detection and localisation of signals, along with moderately improved sensitivity and accuracy over coarsened banks at the same level of computational cost. This is potentially useful for any search involving template banks, and especially in the analysis of data from future space-based detectors such as eLISA, for which online grid searches are difficult due to the long-duration waveforms and large parameter spaces.
\end{abstract}

\maketitle

\section{Introduction}

Advanced LIGO \cite{H2010} has recently made the first direct detection of gravitational waves (GWs) from an astrophysical source \cite{AEA2016}; more detections are now expected routinely from the ground-based interferometer network comprising Advanced LIGO and Advanced Virgo \cite{AEA2011}. These should be followed over the next two decades by detections of nanohertz GW sources using pulsar timing arrays \cite{HEA2010}, and of millihertz sources by the proposed space-based detector eLISA \cite{AEA2012} or more ambitious missions such as DECIGO \cite{KEA2011}. The search for GW signals in noisy data from such detectors---and the follow-up estimation of their source parameters---is contingent upon reliable statistical analysis of the data.

GW signals from sources such as stellar-mass compact binary coalescences or massive black-hole binary inspirals are typically weak compared to the detector noise in which they are embedded. The standard approach in GW data analysis is to correlate the detector data with a bank of waveform templates sampled from the parameter space of a waveform model, which allows signal-to-noise ratio (SNR) to be built up over the detector bandwidth. This correlation is essentially an inner product on the function space of finite-length time series; it must be evaluated numerically for each template, and carries the bulk of the computational cost in online GW searches \cite{JK2009,JK2012}.

Various strategies exist to reduce the online cost of evaluating inner products for GW detection and parameter estimation purposes, typically by shifting the computational burden to the preparatory offline stage. Some methods focus on making individual inner products computationally cheaper: this may be achieved across regions of parameter space through direct interpolation \cite{MDF2005,SEA2013}, or more generally by using a reduced order quadrature \cite{CFGT2013,CEA2015}. Other methods seek to reduce the number of required inner products: either by accelerating the convergence to correlation maxima in a stochastic search \cite{CC2005,CCR2006,FGHP2009}, or through reduced-basis decomposition of the template bank in a grid search \cite{H2009,CEA2010,FEA2011}.

In a recently proposed method for evaluating fewer inner products in a grid search, binary labelling is used to define a compressed non-orthogonal basis that maximises compression losslessly (in the sense of perfect signal recovery without noise) \cite{W2014}. This idea is fully general and admits a much higher compression rate than existing methods based on the eigenvalue structure of the template bank, but comes with significant penalties to detection sensitivity and identification accuracy in the presence of detector noise. The method as originally described also suffers from an arbitrarily asymmetric treatment of templates, as well as a restrictive level of compression that limits its practicality to high-SNR signals. While the binary labelling method might be useful in the context of eLISA (where source SNRs are potentially higher than for ground-based detectors), its practical applicability to GW data analysis remains undeveloped and hence unclear.

In this paper, we introduce and develop the related method of conic compression (i.e. defining a compressed basis through conic combinations of templates) by characterising its performance under various simplifying assumptions, before investigating its viability for current and future GW detectors with a more realistic example. We propose several compression schemes, one of which subsumes a symmetric-treatment version of the binary labelling method \cite{W2014} as a particular case. These tunable schemes feature discrete transitions between zero and maximal compression, and offer fast detection and localisation of GW signals in the search space with a controlled loss (if at all) in sensitivity/accuracy. Their generality and straightforward implementation also allow them to supplement existing grid-search methods, or to rapidly identify seed points for stochastic searches.

The general method of conic compression is set out in Sec.\,\ref{sec:schemes}. Three families of conic compression schemes are then proposed in Secs\,\ref{subsec:partition}--\ref{subsec:binomial}: a lossy scheme based on partitions of the template bank, and two lossless schemes whose conic combinations are determined permutatively or by base representations of template labels. We calculate the optimal detection statistics for these schemes, and find that the standard maximum-overlap statistic is significantly suboptimal for detection in the lossless case. Sec.\,\ref{subsec:comparison} compares the three schemes under simplified conditions, i.e. assuming the GW signal is proportional to a single template in an orthogonal template bank. The lossy partition scheme is shown to have slightly better detection sensitivity than its lossless counterparts at the same level of compression. Furthermore, while the lossless schemes provide automatic identification (i.e. localisation to a single template) of the signal upon detection, the identification accuracy falls off more rapidly with compression in the presence of noise.

We focus exclusively on the partition scheme in Sec.\,\ref{sec:assumptions}, where the orthogonality and single-template assumptions are lifted separately. As shown in Sec.\,\ref{subsec:nonorthogonality}, the overall performance of the scheme is partition-dependent in the case of a correlated (non-orthogonal) template bank, and must be pre-optimised by grouping highly correlated templates together. The optimised partition scheme retains the benefits of a correlated template bank up to high levels of compression, and is superior to a simple ``coarsening'' of the template bank (obtained by increasing the maximal mismatch between neighbouring templates). Sec.\,\ref{subsec:2Dsubspace} discusses the case of a GW signal lying in a low-dimensional subspace of an orthogonal template bank, for which the detection sensitivity of the scheme is not significantly reduced.

In Sec.\,\ref{sec:example}, we implement the optimised partition scheme for a highly correlated (maximal mismatch $\approx0.01$) template bank of $\sim10^4$ post-Newtonian (PN) waveforms, which describe the gravitational radiation emitted during the inspiral phase of a comparable-mass binary merger. The scheme is shown to be viable for practical applications, as it performs well on this example up to high levels of compression and at all considered values of SNR. Its detection rate for a signal injected centrally is superior to that of the coarsening approach (especially at compression rates of over $80\%$), and this improvement is even more marked for a signal injected at the boundary of the bank. In addition, the accuracy rate for localisation of the injection to a $<0.1\%$ region of the search space is undiminished up to a compression level of $90\%$, and is again higher than that of the coarsening approach.

The considerable speed-up and enhanced accuracy in localising the GW signal with conic compression is particularly promising for eLISA data analysis, where the online use of template banks is made challenging by the large parameter spaces of typical sources \cite{GEA2004}. While the long duration of eLISA signals is computationally prohibitive to fully coherent searches even with compression, our method is suitable for the shorter semi-coherent searches that are required for rapid electromagnetic follow-up.

Conic compression might also provide a viable alternative to the singular-value-decomposition (SVD) method used in LIGO detection pipelines for compact binary coalescences \cite{CEA2010}: it scales well with parameter-space dimensionality and easily matches or surpasses the order-of-magnitude computational savings of the SVD method, with any loss of SNR coming mainly from the maximal mismatch of the original template bank (rather than an SVD reconstruction). Furthermore, our method may in principle be used to further compress the reduced bases obtained through the various orthogonal-decomposition methods \cite{H2009,CEA2010,FEA2011}. Whether any computational benefits might be gained from such a combination of the two approaches is left for future investigation.

\section{Compression schemes}\label{sec:schemes}

In the standard GW data analysis framework, data from a detector may be written as the time series
\begin{equation}\label{eq:data}
\mathcal{X}(t)=\mathcal{S}(t)+\mathcal{N}(t),
\end{equation}
where the GW signal $\mathcal{S}(t)$ is a deterministic function of time (and some unknown source parameters), and the additive detector noise $\mathcal{N}(t)$ is a Gaussian and stationary stochastic process.

Matched filtering involves passing the data through some GW template filter $\mathcal{F}(t)$ via convolution, which defines an inner product on the function space of finite-length time series \cite{CF1994}. This inner product is given by
\begin{equation}\label{eq:innerproduct}
\langle\mathcal{X}|\mathcal{F}\rangle=\int_{-\infty}^\infty\frac{\tilde{\mathcal{X}}(f)\tilde{\mathcal{F}}^*(f)}{S_\mathcal{N}(f)}\,df,
\end{equation}
where $S_\mathcal{N}(f)$ is the two-sided spectral density of the detector noise. Since $\mathcal{N}(t)$ is stationary, $S_\mathcal{N}(f)$ is simply the Fourier transform of the autocorrelation function $R_\mathcal{N}(\tau)=\mathrm{E}(\mathcal{N}(t)\mathcal{N}(t-\tau))$, and we have the identity
\begin{equation}
\mathrm{E}(\langle\mathcal{N}|\mathcal{F}\rangle\langle\mathcal{N}|\mathcal{F}'\rangle)=\langle\mathcal{F}|\mathcal{F}'\rangle.
\end{equation}
The SNR $\rho_\mathcal{F}$ of the filtered data is then related to the true SNR $\rho$ by
\begin{equation}\label{eq:SNR}
\rho_\mathcal{F}^2=\frac{\langle\mathcal{S}|\mathcal{F}\rangle^2}{\langle\mathcal{F}|\mathcal{F}\rangle}\leq\langle\mathcal{S}|\mathcal{S}\rangle=\rho^2.
\end{equation}

We now consider a generic bank of $N$ GW templates $h_n(t)$, where the template labels $n$ are drawn from the collection $\mathbf{N}:=\{n\in\mathbb{Z}^+\,|\,n\leq N\}$, and the templates have been normalised such that $\langle h_n|h_n\rangle=1$ for all $n\in\mathbf{N}$. The inner products of the data \eqref{eq:data} and the templates define $N$ associated statistics
\begin{equation}\label{eq:originalstatistic}
x_n:=\langle\mathcal{X}|h_n\rangle,
\end{equation}
which may be used for detection and localisation in a simple grid search.

Our general method of compression is to reduce the number of statistic evaluations from $N$ to $M$ by considering conic (i.e. positive-coefficient) combinations of the original templates. The template labels are grouped into $M$ sets $\mathbf{U}_m$, where the set labels $m$ are drawn from the collection $\mathbf{M}:=\{m\in\mathbb{Z}^+\,|\,m\leq M\}$, and the sets satisfy $\bigcup_{m\in\mathbf{M}}\mathbf{U}_m=\mathbf{N}$. These sets define $M$ conic templates
\begin{equation}\label{eq:conictemplate}
H_m(t):=\sum_{n\in\mathbf{U}_m}h_n(t),
\end{equation}
which are prepared at the offline stage (like the template bank itself), along with $M$ associated statistics
\begin{equation}\label{eq:conicstatistic}
X_m:=\langle\mathcal{X}|H_m\rangle=\sum_{n\in\mathbf{U}_m}x_n,
\end{equation}
which are evaluated at the online stage.

Without any prior assumptions on the template bank, each template must be treated equally. This is done by ensuring that:
\begin{enumerate}[label=(\alph*)]
\item each combination is weighted equally;
\item each combination includes the same number of templates;
\item each template is included in the same number of combinations.
\end{enumerate}
Definition \eqref{eq:conictemplate} has been chosen to satisfy condition (a), while condition (b) is imposed by further requiring $\mathrm{card}(\mathbf{U}_m)=\mathrm{card}(\mathbf{U}_{m'})$ for all $m,m'\in\mathbf{M}$ (where the set cardinality $\mathrm{card}(\mathbf{S})$ is the number of elements in the set $\mathbf{S}$). Condition (c) must be enforced separately in the construction of the sets. The second equality in Definition \eqref{eq:conicstatistic} relates the conic statistic evaluations to the original statistics \eqref{eq:originalstatistic}, which are no longer evaluated at the online stage.

To simplify analysis, we first assume the template bank is an orthogonal set such that
\begin{equation}\label{eq:orthogonality}
\langle h_n|h_{n'}\rangle=\delta_{nn'},
\end{equation}
where $\delta_{ij}$ is the Kronecker delta. We further assume the GW signal (if present) lies in the one-dimensional subspace spanned by a single template in Hilbert space, i.e.
\begin{equation}\label{eq:1Dsubspace}
\mathcal{S}(t)=Ah_1(t),
\end{equation}
where $A>0$ and the templates have been relabelled without loss of generality. It follows from \eqref{eq:SNR} and \eqref{eq:1Dsubspace} that $A=\rho$. These orthogonal and 1-D restrictions are neither realistic nor optimal, but facilitate the analytic assessment and comparison of various compression schemes in this section. The overall performance of conic compression is generally improved by the lifting of these assumptions, which we consider in Secs\,\ref{sec:assumptions} and \ref{sec:example}.

In the presence of a GW signal, the expectation values and covariances of the normally distributed original statistics \eqref{eq:originalstatistic} are now given by
\begin{equation}\label{eq:originalexpectation}
\mathrm{E}(x_n)=A\langle h_1|h_n\rangle=A\delta_{1n},
\end{equation}
\begin{equation}\label{eq:originalcovariance}
\mathrm{cov}(x_n,x_{n'})=\langle h_n|h_{n'}\rangle=\delta_{nn'}.
\end{equation}
As the labelling of templates is itself a probabilistic process with discrete uniform distribution, the original statistic vector $x$ has the multivariate Gaussian distribution $\mathcal{G}(\mu^{(i)},\Sigma)$ (with $\mu^{(i)}_n=\mathrm{E}(x_n)$ and $\Sigma_{nn'}=\mathrm{cov}(x_n,x_{n'})$), but summed over the $N$ possible assignments $i$ of $1\in\mathbf{N}$ and renormalised accordingly. If the signal is absent, the distribution of $x$ is simply $\mathcal{G}(0,\Sigma)$. Hence we have

\begin{widetext}
\begin{equation}\label{eq:vectorp1}
p_1(x)\propto\frac{1}{N}\sum_{i=1}^N\exp{\left(-\frac{1}{2}x^T\Sigma^{-1}x+\mu_{(i)}^T\Sigma^{-1}x-\frac{1}{2}\mu_{(i)}^T\Sigma^{-1}\mu^{(i)}\right)},
\end{equation}
\begin{equation}\label{eq:vectorp0}
p_0(x)\propto\exp{\left(-\frac{1}{2}x^T\Sigma^{-1}x\right)},
\end{equation}
\end{widetext}

\noindent where $p_1$ and $p_0$ are the probability density functions of $x$ in the respective presence or absence of a GW signal.

An optimal detection region $\mathcal{R}$ in Hilbert space maximises the detection rate $P_D=\int_\mathcal{R}p_1$ subject to a given false alarm rate $P_F=\int_\mathcal{R}p_0$; hence $p_1=\lambda p_0$ on its boundary $\partial\mathcal{R}$ for some Lagrange multiplier $\lambda$. Using \eqref{eq:originalexpectation}--\eqref{eq:vectorp0}, we define the optimal detection statistic
\begin{equation}\label{eq:originaloptimal}
x_\mathrm{opt}:=\frac{p_1(x)}{p_0(x)}=\frac{1}{N}\exp{\left(-\frac{A^2}{2}\right)}\sum_{n\in\mathbf{N}}\exp{\left(Ax_n\right)},
\end{equation}
such that the optimal detection surfaces $\partial\mathcal{R}$ are precisely the level sets of $x_\mathrm{opt}$ parametrised by $\lambda$, and a detection is claimed if $x_\mathrm{opt}$ exceeds the threshold $\lambda_T$ corresponding to some fixed value of $P_F$.

In deriving \eqref{eq:originaloptimal}, we have implicitly assumed a population of GW sources with equal likelihood and known signal amplitude. Eq.\,\eqref{eq:originaloptimal} therefore defines the optimal statistic for detecting events drawn from such a population. For a population of sources that are not equally likely, we need to replace the sum in \eqref{eq:originaloptimal} with a suitably weighted sum. Similarly, for a population with a distribution of amplitudes, we need to marginalise \eqref{eq:originaloptimal} over $A$; in the case of an (improper) uniform prior over the interval $(-\infty,\infty)$, this would give a detection statistic proportional to $\sum_{n\in\mathbf{N}}\exp(x_n^2/2)$.

Any choice of population makes assumptions about the astrophysical distribution of GW sources that might not be justified. In this paper, the focus is on the investigation and comparison of template bank compression schemes, and so we only consider the equal-likelihood and known-amplitude population assumed in the derivation of \eqref{eq:originaloptimal}. While the treatment of amplitude in particular is artificial, a search that is optimised for sensitivity to signal amplitudes around the detection threshold will likely be near-optimal for any given astrophysical population (and closer to optimality than a search tuned for the wrong astrophysical population). Finally, we note that although \eqref{eq:originaloptimal} has been derived as a frequentist optimal statistic, the same equation also arises as the Bayes factor for the presence (versus absence) of a signal, assuming flat model priors and the source population assumptions outlined above.

\begin{figure}
\centering
\includegraphics[width=0.8\columnwidth]{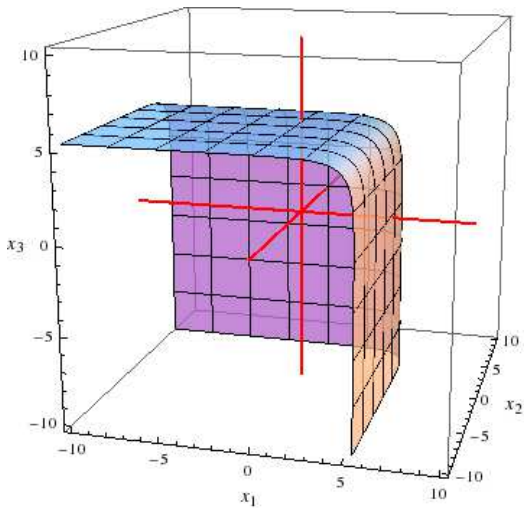}
\caption{Three-dimensional projection of optimal detection surface for uncorrelated statistics $x_n$, at true SNR of 2.}
\label{fig:uncorrelatedsurface}
\end{figure}

For sufficiently high SNR (large $A$), the optimal surfaces $x_\mathrm{opt}=\lambda$ defined by \eqref{eq:originaloptimal} are well approximated by semi-infinite hypercubes in Hilbert space (see Fig.\,\ref{fig:uncorrelatedsurface}), i.e. the level sets of the standard grid-search detection statistic \cite{DS1994,A1995,O1996}
\begin{equation}
x_\mathrm{max}=\max_{n\in\mathbf{N}}\{x_n\}.
\end{equation}
Since the original statistics \eqref{eq:originalstatistic} are uncorrelated, the probability density functions of $x_\mathrm{max}$ in the presence or absence of a GW signal are obtainable explicitly. These are given respectively by
\begin{multline}\label{eq:maximump1}
q_1(x_\mathrm{max})=F_0(x_\mathrm{max})^{N-1}f_1(x_\mathrm{max})\\+(N-1)F_0(x_\mathrm{max})^{N-2}F_1(x_\mathrm{max})f_0(x_\mathrm{max}),
\end{multline}
\begin{equation}\label{eq:maximump0}
q_0(x_\mathrm{max})=NF_0(x_\mathrm{max})^{N-1}f_0(x_\mathrm{max}),
\end{equation}
where $f_s(x_\mathrm{max})$ is the probability density function for the Gaussian distribution $\mathcal{G}(sA,1)$, and $F_s(x_\mathrm{max})$ is the cumulative distribution function
\begin{equation}
F_s(x_\mathrm{max})=\int_{-\infty}^{x_\mathrm{max}}f_s(u)\,du.
\end{equation}

For our analysis of conic compression schemes, we also require the expectation values and covariances of the normally distributed conic statistics \eqref{eq:conicstatistic}. From \eqref{eq:conicstatistic}, \eqref{eq:originalexpectation} and \eqref{eq:originalcovariance}, it follows in the presence of a GW signal that
\begin{equation}\label{eq:conicexpectation}
\mathrm{E}(X_m)=\sum_{n\in\mathbf{U}_m}\mathrm{E}(x_n)=A\,\mathrm{card}(\{1\}\cap\mathbf{U}_m),
\end{equation}
\begin{multline}\label{eq:coniccovariance}
\mathrm{cov}(X_m,X_{m'})=\sum_{n\in\mathbf{U}_m}\sum_{n'\in\mathbf{U}_{m'}}\mathrm{cov}(x_n,x_{n'})\\=\mathrm{card}(\mathbf{U}_m\cap\mathbf{U}_{m'}),
\end{multline}
where the cardinalities are determined by the choice of compression scheme. As before, the conic statistic vector $X$ has the multivariate Gaussian distribution $\mathcal{G}(\mu^{(i)},\Sigma)$ (now with $\mu^{(i)}_m=\mathrm{E}(X_m)$ and $\Sigma_{mm'}=\mathrm{cov}(X_m,X_{m'})$), but summed over the $N$ possible assignments of $1\in\mathbf{N}$ and renormalised accordingly. If the signal is absent, the distribution of $X$ is again $\mathcal{G}(0,\Sigma)$. The probability density functions of $X$ in the presence or absence of a GW signal are then given respectively by \eqref{eq:vectorp1} and \eqref{eq:vectorp0} with $x\equiv X$.

We now propose and investigate three general conic compression schemes in Secs\,\ref{subsec:partition}--\ref{subsec:binomial}, before comparing their performance and potential applicability in Sec.\,\ref{subsec:comparison}. The orthogonal and 1-D restrictions \eqref{eq:orthogonality} and \eqref{eq:1Dsubspace} are assumed throughout Sec.\,\ref{sec:schemes}.

\subsection{Partition scheme}\label{subsec:partition}

The simplest method of grouping the template labels $n$ is to take the family of sets $\mathbf{U}_m$ as a partition of $\mathbf{N}$, i.e. $\mathbf{U}_m\cap\mathbf{U}_{m'}=\emptyset$ for all distinct $m,m'\in\mathbf{M}$. Condition (c) is then automatically satisfied, while condition (b) defines the set cardinality $P=\mathrm{card}(\mathbf{U}_m)$ for all $m\in\mathbf{M}$. It follows that $M=N/P$.

For the comparison of schemes in Sec.\,\ref{subsec:comparison}, it is useful to introduce a compression parameter $K\in\mathbb{Z}^+$ for each scheme, which determines the compression rate
\begin{equation}\label{eq:compression}
\kappa:=1-\frac{N_\mathrm{eval}}{N},
\end{equation}
where $N_\mathrm{eval}=M$ is the required number of statistic evaluations (for detection or localisation purposes). This generates a sliding scale of groupings that ranges from no compression at $K=1$ to maximal compression at some scheme-dependent value of $K$. We may clearly choose $K=P$ for the partition scheme, such that maximal compression is given by $K=N$. The minimal nontrivial compression is $50\%$ at $K=2$, while there are diminishing returns at large $K$ since $\kappa(K)$ is concave-down.

From \eqref{eq:conicexpectation} and \eqref{eq:coniccovariance}, we now have
\begin{equation}
\mathrm{E}(X_m)=A\delta_{1m},
\end{equation}
\begin{equation}
\mathrm{cov}(X_m,X_{m'})=P\delta_{mm'},
\end{equation}
where the sets have been relabelled such that $1\in\mathbf{U}_1$ without loss of generality. Again considering the $N$ possible assignments of $1\in\mathbf{N}$, the optimal detection statistic $X_\mathrm{opt}:=p_1(X)/p_0(X)$ follows from \eqref{eq:vectorp1} and \eqref{eq:vectorp0} (with $x\equiv X$) as
\begin{equation}\label{eq:partitionoptimal}
X_\mathrm{opt}=\frac{1}{M}\exp{\left(-\frac{A^2}{2P}\right)}\sum_{m\in\mathbf{M}}\exp{\left(\frac{A}{P}X_m\right)}.
\end{equation}

\begin{figure}
\centering
\includegraphics[width=0.9\columnwidth]{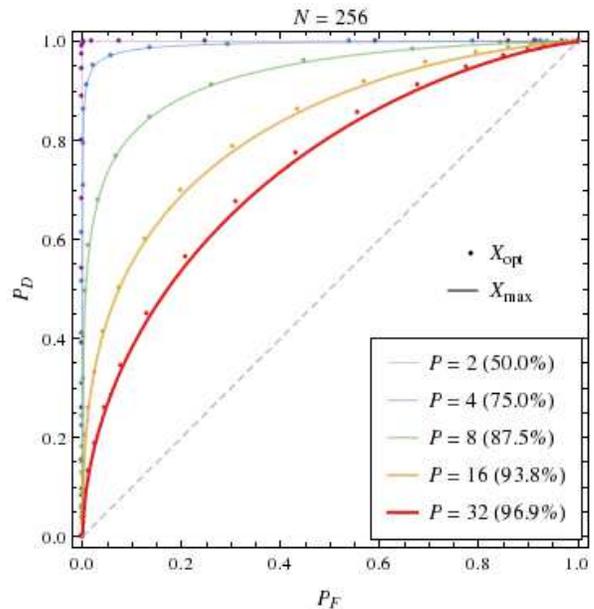}
\caption{ROC curves for the partition scheme's optimal and maximum-overlap detection statistics, at different values of set cardinality $P$ (with compression rate $\kappa$ in parentheses) for a 256-template bank and a true SNR of 10. The dashed diagonal line indicates the worst possible performance, i.e. a random search for which the detection and false alarm rates are equal.}
\label{fig:partitionROC}
\end{figure}

Since the conic statistics for the partition scheme remain uncorrelated, the optimal surfaces $X_\mathrm{opt}=\lambda$ resemble that in Fig.\,\ref{fig:uncorrelatedsurface}, and in lieu of \eqref{eq:partitionoptimal} it is valid to consider the maximum-overlap detection statistic
\begin{equation}\label{eq:partitionmaximum}
X_\mathrm{max}=\max_{m\in\mathbf{M}}\{X_m\}.
\end{equation}
Receiver operating characteristic (ROC) curves of detection rate $P_D$ against false alarm rate $P_F$ for both the optimal and maximum-overlap statistics are compared in Fig.\,\ref{fig:partitionROC}.\footnote{The curves for \eqref{eq:partitionoptimal} were obtained via $10^5$-trial Monte Carlo simulations, while numerical integration of \eqref{eq:maximump1} and \eqref{eq:maximump0} was used to generate quicker and more precise curves for \eqref{eq:partitionmaximum}.} With increased compression, the performance of the maximum-overlap statistic falls away slightly from that of the optimal statistic, due to the lowering of effective SNR $A/\sqrt{P}$ in \eqref{eq:partitionoptimal}; nevertheless, \eqref{eq:partitionmaximum} is a sound approximation as both sets of ROC curves show good overall agreement.

For the partition scheme to admit a useful (i.e. populated) sliding scale of compression rates, the template bank might need to be trimmed or padded such that $N$ has as many divisors as possible. Fixing the false alarm rate and choosing either a desired detection rate or a compression rate then allows advance determination of the conic templates \eqref{eq:conictemplate} and the threshold $\lambda_T$, which is the value of $\lambda$ corresponding to the fixed false alarm rate. The algorithm for GW detection follows as: (i) evaluate the conic statistics \eqref{eq:conicstatistic}; (ii) claim a detection if $X_\mathrm{max}>\lambda_T$. Threshold and detection SNRs for the maximum-overlap statistic are defined respectively as
\begin{equation}
\rho_T:=\frac{\lambda_T}{\sqrt{\mathrm{var}(X_\mathrm{max})}},
\end{equation}
\begin{equation}
\rho_D:=\frac{X_\mathrm{max}}{\sqrt{\mathrm{var}(X_\mathrm{max})}}.
\end{equation}

An extension of the detection algorithm is required for identification purposes (i.e. localisation to a single template), since the simple coarse-graining of partition compression does not distinguish between template labels in the same set. The signal is most likely to be associated with the largest conic statistic evaluation $X_{(1)}$, so the best candidate template may be obtained by further evaluating all the original statistics $x_n$ contributing to $X_{(1)}$ and identifying the largest. This finer level of evaluations increases the computational cost by $P$ to $N_\mathrm{eval}=M+P$.

For better identification accuracy at lower SNRs, we may widen our search to the $i$ largest $X_m$ instead, at an added computational cost of $iP$. The standard algorithms $\mathrm{I}_i$ for GW identification follow (after detection) as: (iii) evaluate the original statistics \eqref{eq:originalstatistic} for all $n\in\mathbf{V}_i$, where
\begin{equation}
\mathbf{V}_i=\bigcup_{j=1}^i\mathbf{U}_{(j)},
\end{equation}
with $\mathbf{U}_{(j)}$ corresponding to the $j$-th largest conic statistic evaluation; (iv) identify $\max_{n\in\mathbf{V}_i}\{x_n\}$.

Other identification algorithms may also be considered. One such alternative is obtained by defining a further partition of $\mathbf{V}_i$ into two sets and evaluating the associated conic statistics, then identifying the set $\mathbf{V}'_i$ corresponding to the larger statistic evaluation and repeating the process with $\mathbf{V}_i\equiv\mathbf{V}'_i$ until $\mathrm{card}(\mathbf{V}'_i)=1$. This method might be useful for large values of $P$; it yields a smaller added computational cost of $2\log_2iP$, but incurs a penalty to identification accuracy since the early iterations still involve coarse-grained searches.

\subsection{Symmetric base scheme}\label{subsec:base}

\begin{figure*}
\centering
\captionsetup[subfloat]{position=top}
\subfloat[]{\includegraphics[width=0.8\columnwidth]{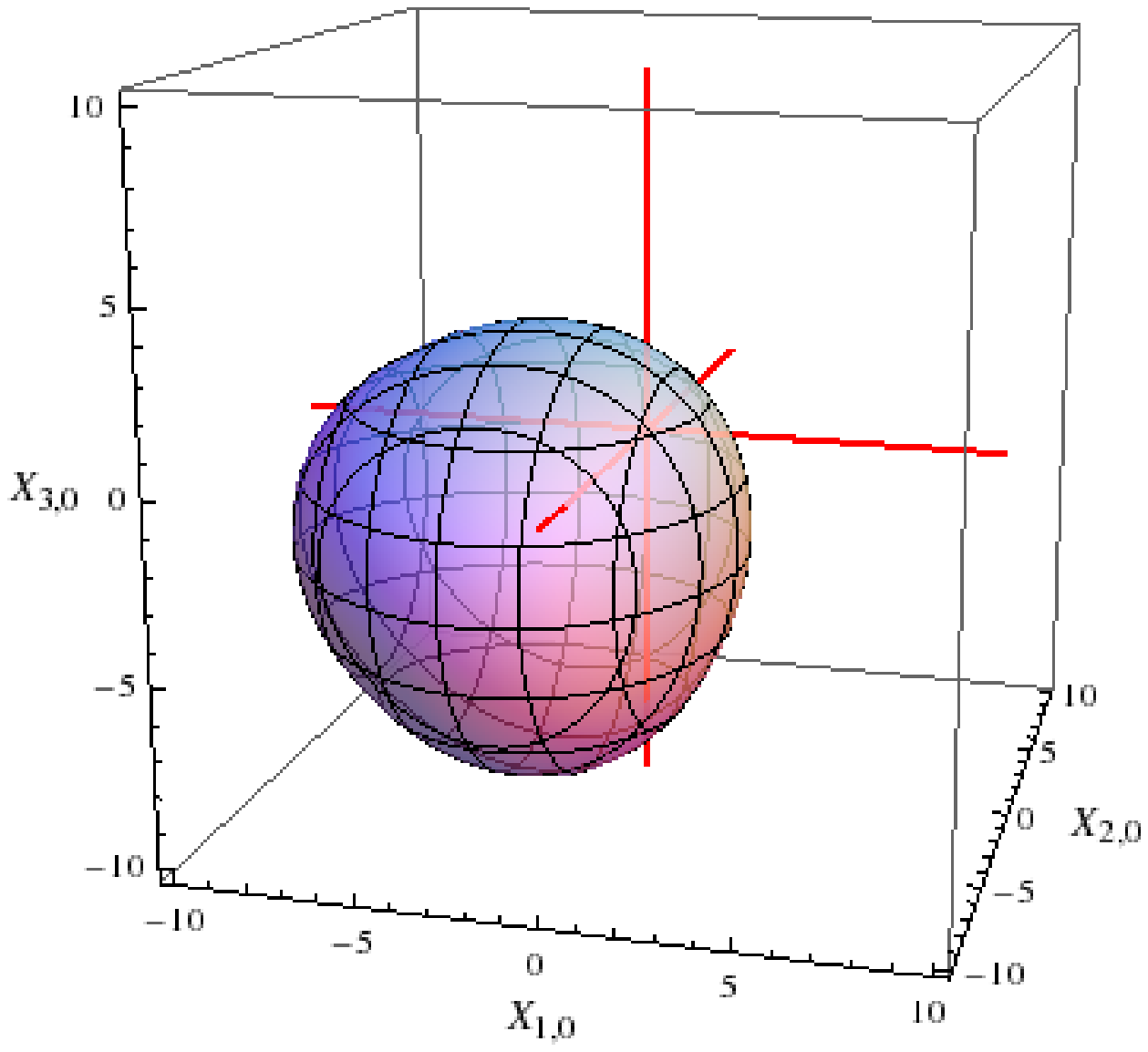}}
\subfloat[]{\includegraphics[width=0.8\columnwidth]{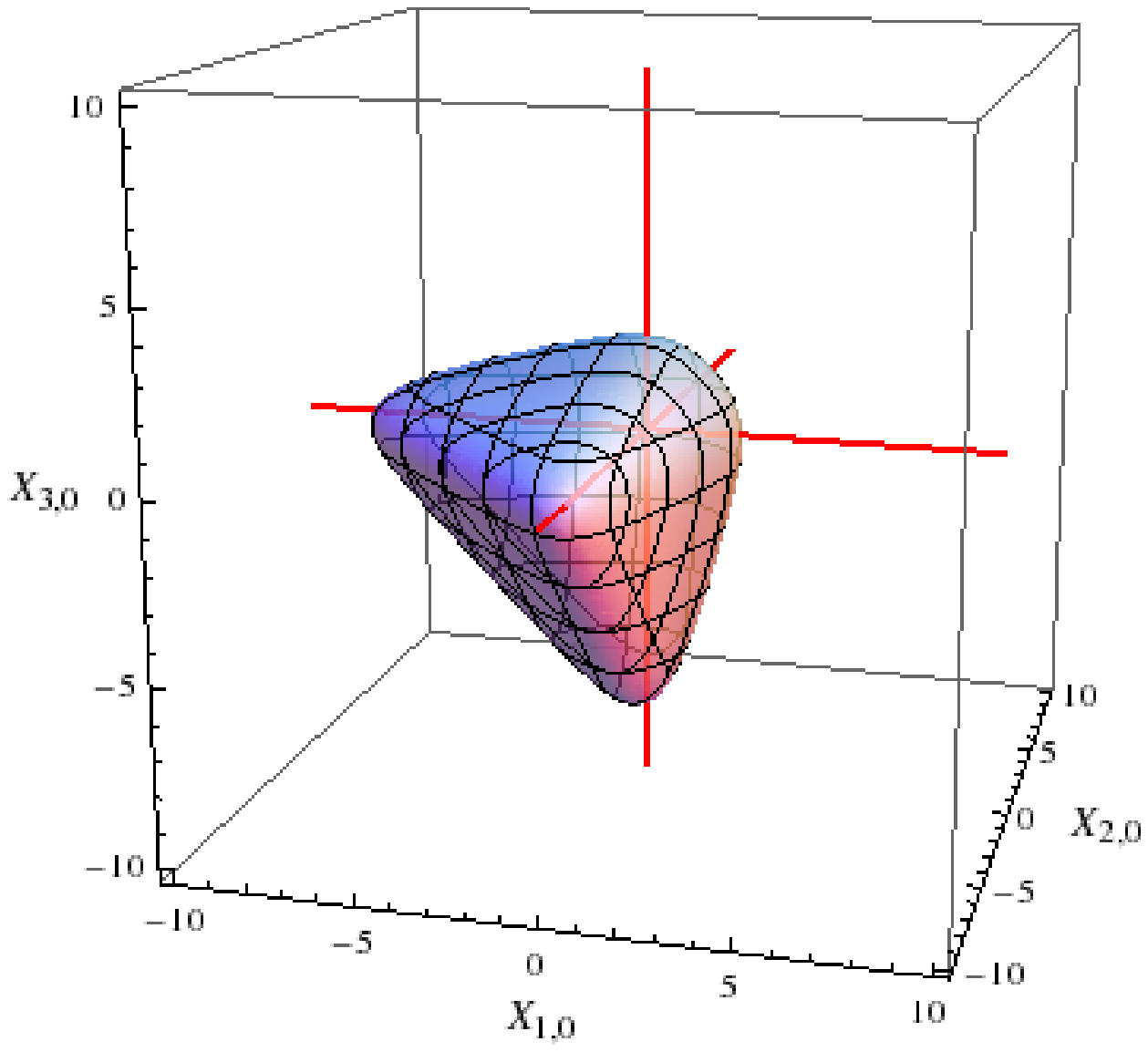}}
\caption{Three-dimensional projection of optimal detection surface for correlated statistics $X_{k,b}$ of symmetric base scheme with $N=256$ and $B=4$, at true SNRs of (a) 10 and (b) 100.}
\label{fig:basesurface}
\end{figure*}

\begin{figure}
\centering
\includegraphics[width=0.9\columnwidth]{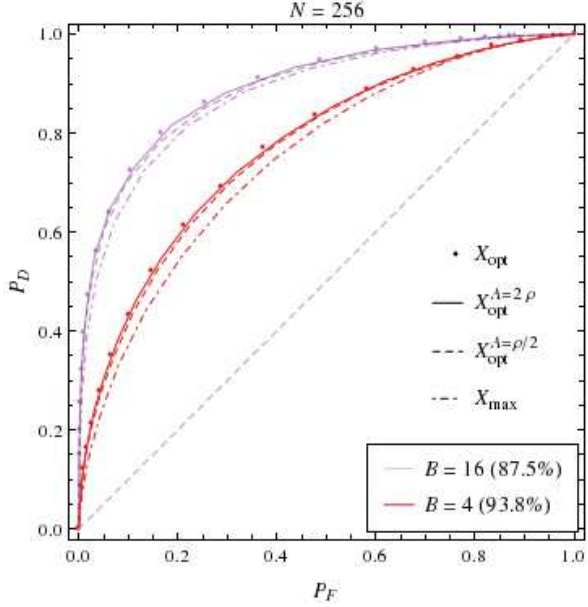}
\caption{ROC curves for the symmetric base scheme's optimal, maximum-overlap and estimated detection statistics, at different values of base $B$ (with compression rate $\kappa$ in parentheses) for a 256-template bank and a true SNR of 10.}
\label{fig:baseROC}
\end{figure}

Without an additional fine-grained search, partition compression is lossy in the sense that the GW signal is not automatically identified in the limit of zero noise. A recently proposed conic compression scheme introduces a lossless method of compression, by representing each template label $n$ in binary and assigning it to the set $\mathbf{U}_m$ if its $m$-th digit is 1 \cite{W2014}. This binary scheme features the largest possible lossless compression ($M=\log_2N$) and an automatic identification of the GW signal; however, it suffers from an unequal treatment of templates (i.e. it violates conditions (b) and (c)) and hence it yields an arbitrary level of performance that depends on the initial assignment of template labels. Furthermore, the restriction to maximum compression limits its usefulness in practical applications.

We propose a compression scheme modelled on the binary labelling method, but symmetrised (for equal treatment of templates) and generalised to a sliding scale of base representations (for tunable compression). The template labels $n$ are represented modulo $N$ in base $B$, and each set $\mathbf{U}_m\equiv\mathbf{U}_{k,b}$ is constructed by collecting all the labels whose $k$-th digit is $b$ (this includes $b=0$, and gives a symmetric version of the binary scheme \cite{W2014} when $B=2$). For conditions (b) and (c) to be satisfied, we require $\log_BN\in\mathbb{Z}^+$; it follows that $M=B\log_BN$.

The compression parameter is chosen as $K=\log_BN$, such that maximal compression is given by $K=\log_3N\approx\ln N$ (base-2 compression is slightly suboptimal with symmetrisation). In contrast to the partition scheme, compression for the symmetric base scheme is dependent on the size of the template bank; the minimal nontrivial compression for $N=10^2$ is nearly $80\%$ at $K=2$ (base-$\sqrt{N}$ compression), and over $95\%$ for $N=10^4$.

From \eqref{eq:conicexpectation} and \eqref{eq:coniccovariance}, we have
\begin{equation}
\mathrm{E}(X_m)=A\delta_{0b},
\end{equation}
\begin{equation}\label{eq:basecovariance}
\mathrm{cov}(X_m,X_{m'})=B^{K-2}(B\delta_{kk'}\delta_{bb'}-\delta_{kk'}+1),
\end{equation}
where $m=B(k-1)+b+1$, and the templates have been relabelled such that $\mathcal{S}(t)=Ah_N(t)$ without loss of generality.\footnote{The covariance matrix defined by \eqref{eq:basecovariance} is rank-deficient, but we may take the Moore--Penrose pseudoinverse $\Sigma^+$ as a suitable (perturbative) approximation to $\Sigma^{-1}$ in \eqref{eq:vectorp1} and \eqref{eq:vectorp0}.} Considering the $N$ possible assignments of $N\in\mathbf{N}$, the optimal detection statistic follows from \eqref{eq:vectorp1} and \eqref{eq:vectorp0} as
\begin{multline}\label{eq:baseoptimal}
X_\mathrm{opt}=\frac{1}{N}\exp{\left(-\frac{\beta KA^2}{2}+(\beta-\alpha)A\,\mathrm{tr}{(X)}\right)}\\\times\prod_{k=1}^K\sum_{b=0}^{B-1}\exp{\left(\alpha AX_{k,b}\right)},
\end{multline}
\begin{equation}
\alpha=\frac{B}{N},\quad\beta=\frac{M-K+1}{NK},
\end{equation}
where $\mathrm{tr}{(X)}:=\sum_{m\in\mathbf{M}}X_m$.

The higher compression rates provided by the symmetric base scheme result from the non-empty intersections among the sets $\mathbf{U}_{k,b}$ with different values of $k$. As seen in \eqref{eq:basecovariance}, these also lead to correlations among the conic statistics $X_{k,b}$. The optimal detection surfaces given by $X_\mathrm{opt}=\lambda$ differ significantly from that depicted in Fig.\,\ref{fig:uncorrelatedsurface}; their projections onto the correlated subspaces are now compact hyperboloids, and no longer approach the semi-infinite hypercubes of the maximum-overlap detection statistic at high SNR (see Fig.\,\ref{fig:basesurface}).

Without a simple approximation for the optimal detection statistic, the most feasible option is to use \eqref{eq:baseoptimal} itself with an estimate $A$ of the true SNR. ROC curves for the estimated statistic $X_\mathrm{opt}^{A=\epsilon\rho}$ with $\epsilon\in\{1/2,2\}$ are compared against those for the optimal and maximum-overlap statistics in Fig.\,\ref{fig:baseROC}. Not much detection sensitivity (for a fixed false alarm rate) is lost if the true SNR can be estimated to within a factor of two, while usage of the maximum-overlap statistic now incurs a more noticeable drop in performance as expected.

The restriction of $N$, $B$ and $K$ to integer values also results in more sparsely populated sliding scales than those admitted by the partition scheme. There are two possible compression rates for $N=256$ (base-2 compression is suboptimal compared to $B=4$), and three for $N=6561=81^2=9^4=3^8$; most other values of $N$ will admit only one or none. Notwithstanding the lack of tunability, a feasible strategy is to trim or pad the template bank such that $N$ is a perfect square or cube, since the smallest values of $K$ already yield high compression rates. The GW detection algorithm then follows as given in Sec.\,\ref{subsec:partition}, with some estimated detection statistic $X_\mathrm{opt}^{A=\epsilon\rho}$ in place of $X_\mathrm{max}$.

One key feature of the symmetric base scheme and other lossless methods of compression is automatic identification of the GW signal (upon detection). In this case, the label of the identified template in base-$B$ representation is given digit-wise by the largest conic statistic evaluation $X_{k,(1)}:=\max_b\{X_{k,b}\}$ for each value of $k$. However, as each digit $k$ is identified individually, the overall identification accuracy falls off severely with increasing $K$ (i.e. the total number of digits).

A possible modification for better accuracy is to consider the $i+1$ largest $X_{k,b}$ for each $k$ and perform an additional fine-grained search over the $(i+1)^K$ templates, which increases the computational cost accordingly. The standard GW identification algorithms $\mathrm{I}_i$ follow (after detection) as: (iii) evaluate the original statistics \eqref{eq:originalstatistic} for all $n\in\mathbf{V}_i$, where
\begin{equation}
\mathbf{V}_i=\bigcap_{k=1}^K\bigcup_{j=1}^{i+1}\mathbf{U}_{k,(j)},
\end{equation}
with $\mathbf{U}_{k,(j)}$ corresponding to the $j$-th largest conic statistic evaluation for each $k$; (iv) identify $\max_{n\in\mathbf{V}_i}\{x_n\}$. Automatic identification is recovered for $i=0$, where steps (iii) and (iv) become unnecessary as $\mathrm{card}(\mathbf{V}_0)=1$.

For large values of $K$ (small values of $B$), the standard identification algorithms might still suffer from poor accuracy. One alternative algorithm is obtained by defining some threshold $X_T$ and considering all conic statistic evaluations $X_{k,b}\geq X_T$, then performing the additional fine-grained search over all the corresponding templates. Such a threshold may be set prior to data-taking; if $X_{k,(1)}<X_T$ for some value of $k$, the $k$-th digit of the number is unconstrained and templates corresponding to all possible choices of that digit are considered. Alternatively, $X_T$ may be based on the data by setting $X_T=f\min_k\{X_{k,(1)}\}$ for some fixed fraction $f$, which ensures that at least one possible value is identified for each digit. Both approaches will in general yield increased accuracy, but they offer less control over the number of conic statistic evaluations considered and hence the overall computational cost.

\subsection{Binomial coefficient scheme}\label{subsec:binomial}

The symmetric base labelling method is not the only construction of the sets $\mathbf{U}_m$ that preserves both lossless compression (automatic identification) and equal treatment of templates (conditions (b) and (c)). In general, we may represent any assignment of $N$ templates to $M$ sets with a collection of $N$ $M$-digit binary labels, where the $m$-th digit of each label is 1 if it appears in $\mathbf{U}_m$ and 0 otherwise. Condition (c) implies that each label must appear in exactly $R$ sets, and hence contain exactly $R$ 1's. In addition, condition (b) defines the set cardinality $C=\mathrm{card}(\mathbf{U}_m)$ for all $m\in\mathbf{M}$, which yields the constraint $NR=MC$ (each of the $N$ labels appears exactly $R$ times across all sets, while each of the $M$ sets contains exactly $C$ labels). For some given integers $N\geq M\geq R$, this constraint is equivalent to the existence of
\begin{equation}\label{eq:constraintC}
C=\frac{NR}{M}\in\mathbb{Z}^+,
\end{equation}
which is both a necessary and sufficient condition for such a set construction to be possible \cite{B2002}.

We now require that the conic statistics \eqref{eq:conicstatistic} are correlated symmetrically, as seen in the partition scheme (but not the symmetric base scheme). This additional condition implies that the intersection of each pair of sets has fixed cardinality $I$, i.e. $\mathrm{card}(\mathbf{U}_m\cap\mathbf{U}_{m'})=I$ for all distinct $m,m'\in\mathbf{M}$. Considering the family of all such intersections then yields the constraint $NR(R-1)=M(M-1)I$ (each of the $N$ labels appears exactly ${^R}\mathrm{C}_2$ times across all intersections, while each of the ${^M}\mathrm{C}_2$ intersections contains exactly $I$ labels). For some given integers $N\geq M\geq R$ and $C$ satisfying \eqref{eq:constraintC}, this constraint is equivalent to the existence of
\begin{equation}\label{eq:constraintI}
I=\frac{NR(R-1)}{M(M-1)}=\frac{C(R-1)}{M-1}\in\mathbb{Z}^+,
\end{equation}
which is a necessary (but not in general sufficient) condition for such a set construction to be possible.

The general construction of a family of sets under the constraints \eqref{eq:constraintC} and \eqref{eq:constraintI} is an open problem in combinatorial design theory (see App. \ref{app:combinatorics}). In this paper, we restrict our focus to a special case that may be treated in greater detail. Every $M$-digit binary number with exactly $R$ 1's is taken to represent a distinct template label; the set cardinality then equals the number of ($M-1$)-digit binary numbers with exactly ($R-1$) 1's, while the intersection cardinality of each pair of sets equals the number of ($M-2$)-digit binary numbers with exactly ($R-2$) 1's. Hence for all distinct $m,m'\in\mathbf{M}$, we have
\begin{equation}
N={^M}\mathrm{C}_R,\quad C={^{M-1}}\mathrm{C}_{R-1},\quad I={^{M-2}}\mathrm{C}_{R-2},
\end{equation}
such that \eqref{eq:constraintC} and \eqref{eq:constraintI} are satisfied. We refer to this as the binomial coefficient scheme, for obvious reasons. The usual ordering of the binary numbers gives a natural map onto the original label collection $\mathbf{N}=\{n\in\mathbb{Z}^+\,|\,n\leq N\}$, although the inverse map is analytically nontrivial (but straightforward in practice).

As the binomial coefficient scheme shares many similarities with the symmetric base scheme, we only highlight its key features in this section. The compression parameter is chosen as $K=R$, such that maximal compression is given by $K=\mathrm{cbc}^{-1}(N)/2$ (where $\mathrm{cbc}(M):=\Gamma(M+1)/\Gamma(M/2+1)^2$ is the continuous extension of the central binomial coefficient ${^M}\mathrm{C}_{M/2}$). Compression rates again depend on the size of the template bank; at small values of $K$, they are only slightly higher than those of the symmetric base scheme.

From \eqref{eq:conicexpectation} and \eqref{eq:coniccovariance}, we have
\begin{equation}
\mathrm{E}(X_m)=A\sum_{r=1}^R\delta_{rm},
\end{equation}
\begin{equation}\label{eq:binomialcovariance}
\mathrm{cov}(X_m,X_{m'})={^{M-2}}\mathrm{C}_{R-2}\left(\frac{M-R}{R-1}\delta_{mm'}+1\right),
\end{equation}
where the sets have been relabelled such that $1\in\mathbf{U}_r$ for $1\leq r\leq R$ without loss of generality. Considering the $N$ possible assignments of $1\in\mathbf{N}$, the optimal detection statistic follows from \eqref{eq:vectorp1} and \eqref{eq:vectorp0} as
\begin{multline}\label{eq:binomialoptimal}
X_\mathrm{opt}=\frac{1}{N}\exp{\left(-\frac{\beta KA^2}{2}+(\beta-\alpha)A\,\mathrm{tr}{(X)}\right)}\\\times\sum_{i=1}^N\exp{\left(\alpha A\sum_{r\in\mathbf{R}_i}X_r\right)},
\end{multline}
\begin{equation}
\alpha=\frac{1}{{^{M-1}}\mathrm{C}_{R-1}},\quad\beta=\frac{1}{{^{M-2}}\mathrm{C}_{R-1}},
\end{equation}
where the sets $\mathbf{R}_i$ are the $N$ distinct $R$-combinations of the collection $\mathbf{M}$.

All the conic statistics $X_m$ are correlated symmetrically, as seen in \eqref{eq:binomialcovariance}. Upon projection onto any three-dimensional subspace, the optimal detection surfaces given by $X_\mathrm{opt}=\lambda$ resemble those in Fig.\,\ref{fig:basesurface} at both low and high SNR. It follows that the maximum-overlap detection statistic is again an inadequate approximation to the optimal statistic, and we are compelled to use \eqref{eq:binomialoptimal} itself (assuming an accurate estimate of true SNR). We do not include ROC curves for the binomial coefficient scheme here, as they are very similar to those in Fig.\,\ref{fig:baseROC}.

A direct comparison of the base and binomial schemes is difficult, since there are few suitable values of $N$ that are exactly valid for both schemes. Lack of tunability is also more of an issue for the binomial scheme: the only values of $N$ that admit more than one nontrivial compression rate might be the Singmaster numbers (which admit two as they appear six times in Pascal's triangle), and it is not known whether any number admits more than two (apart from $N=3003={^{78}}\mathrm{C}_2={^{15}}\mathrm{C}_5={^{14}}\mathrm{C}_6$) \cite{S1971,S1975}. The problem may be overcome by considering a more general compression scheme satisfying the conditions \eqref{eq:constraintC} and \eqref{eq:constraintI}. This is beyond the scope of the current paper due to the complexity of set construction (see App. \ref{app:combinatorics}), but might be investigated for specific template banks in the future.

The GW detection algorithm for the binomial coefficient scheme is as given in Sec.\,\ref{subsec:partition}, with some estimated detection statistic $X_\mathrm{opt}^{A=\epsilon\rho}$ in place of $X_\mathrm{max}$. Automatic identification is available as well, with the label of the identified template given uniquely by the $R$ largest conic statistic evaluations. For higher accurate-identification rates, a possible alternative is to consider the $R+i$ largest $X_m$ and perform an additional fine-grained search over the ${^{R+i}}\mathrm{C}_R$ templates. The standard GW identification algorithms $\mathrm{I}_i$ follow (after detection) as: (iii) evaluate the original statistics \eqref{eq:originalstatistic} for all $n\in\mathbf{V}_i$, where
\begin{equation}
\mathbf{V}_i=\bigcup_{k=1}^{{^{R+i}}\mathrm{C}_R}\bigcap_{j\in\mathbf{J}_k}\mathbf{U}_{(j)},
\end{equation}
with $\mathbf{U}_{(j)}$ corresponding to the $j$-th largest conic statistic evaluation and the sets $\mathbf{J}_{k}$ given by the ${^{R+i}}\mathrm{C}_R$ distinct $R$-combinations of $\{j\in\mathbb{Z}^+\,|\,j\leq R+i\}$; (iv) identify $\max_{n\in\mathbf{V}_i}\{x_n\}$. Automatic identification is recovered for $i=0$, where steps (iii) and (iv) become unnecessary as $\mathrm{card}(\mathbf{V}_0)=1$.

We note that another possible scheme would be a ``direct sum'' of the partition scheme and either the symmetric base or binomial coefficient scheme. The collection of template labels is first partitioned into subcollections, each of which is further decomposed into smaller sets via one of the correlated schemes; these sets may also be recombined across the initial partition for increased compression. We do not consider this further here, but such an approach would overcome some of the difficulties associated with the restricted values of $N$ for the base and binomial schemes.

\subsection{Performance comparison}\label{subsec:comparison}

\begin{figure*}
\centering
\includegraphics[width=0.8\textwidth]{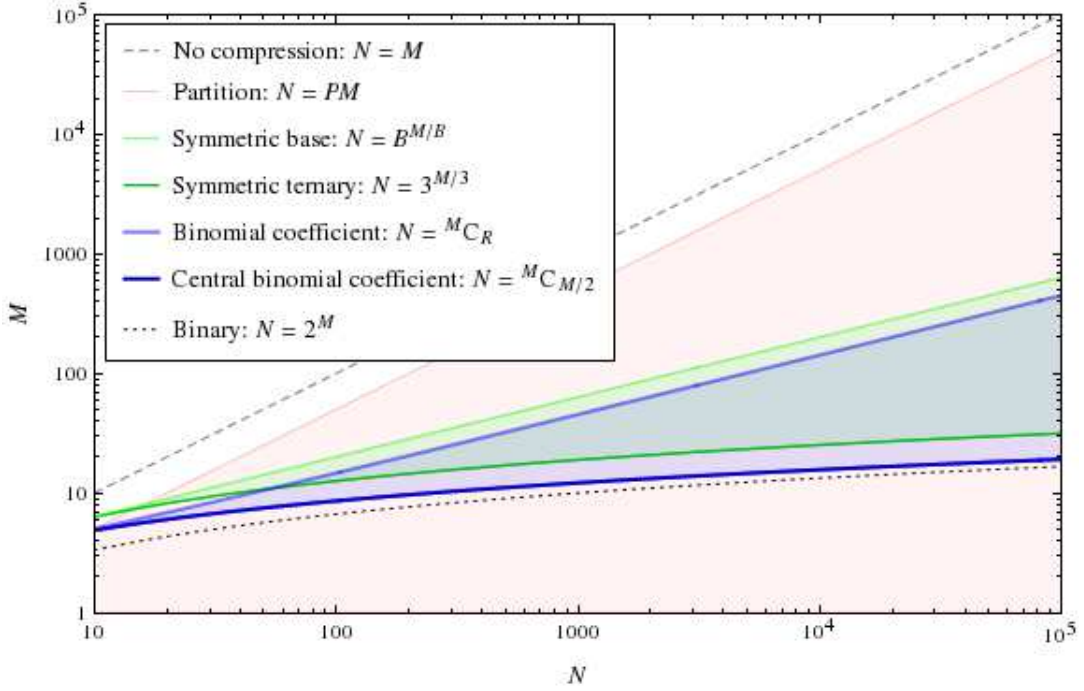}
\caption{Log--log plots of $M$ against $N$ for various compression schemes. For each tunable scheme, the corresponding shaded region indicates the range of possible compression rates (with the trivial compression setting $K=1$ excluded). Not every point in this region is realisable in practice, as discussed in the text.}
\label{fig:comparecompression}
\end{figure*}

\begin{figure*}
\centering
\captionsetup[subfloat]{position=top}
\subfloat[]{\includegraphics[width=0.9\columnwidth]{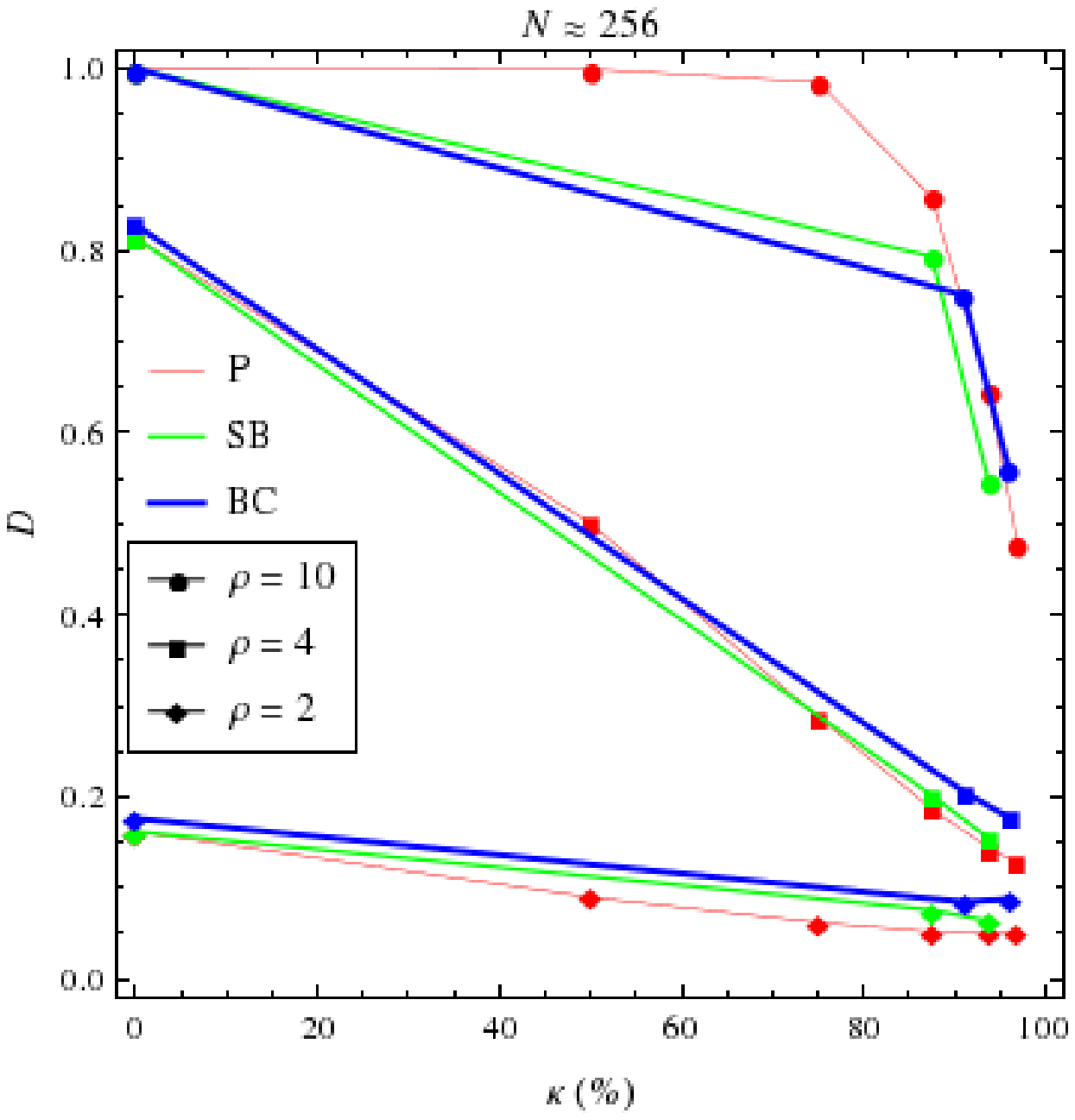}}
\subfloat[]{\includegraphics[width=0.9\columnwidth]{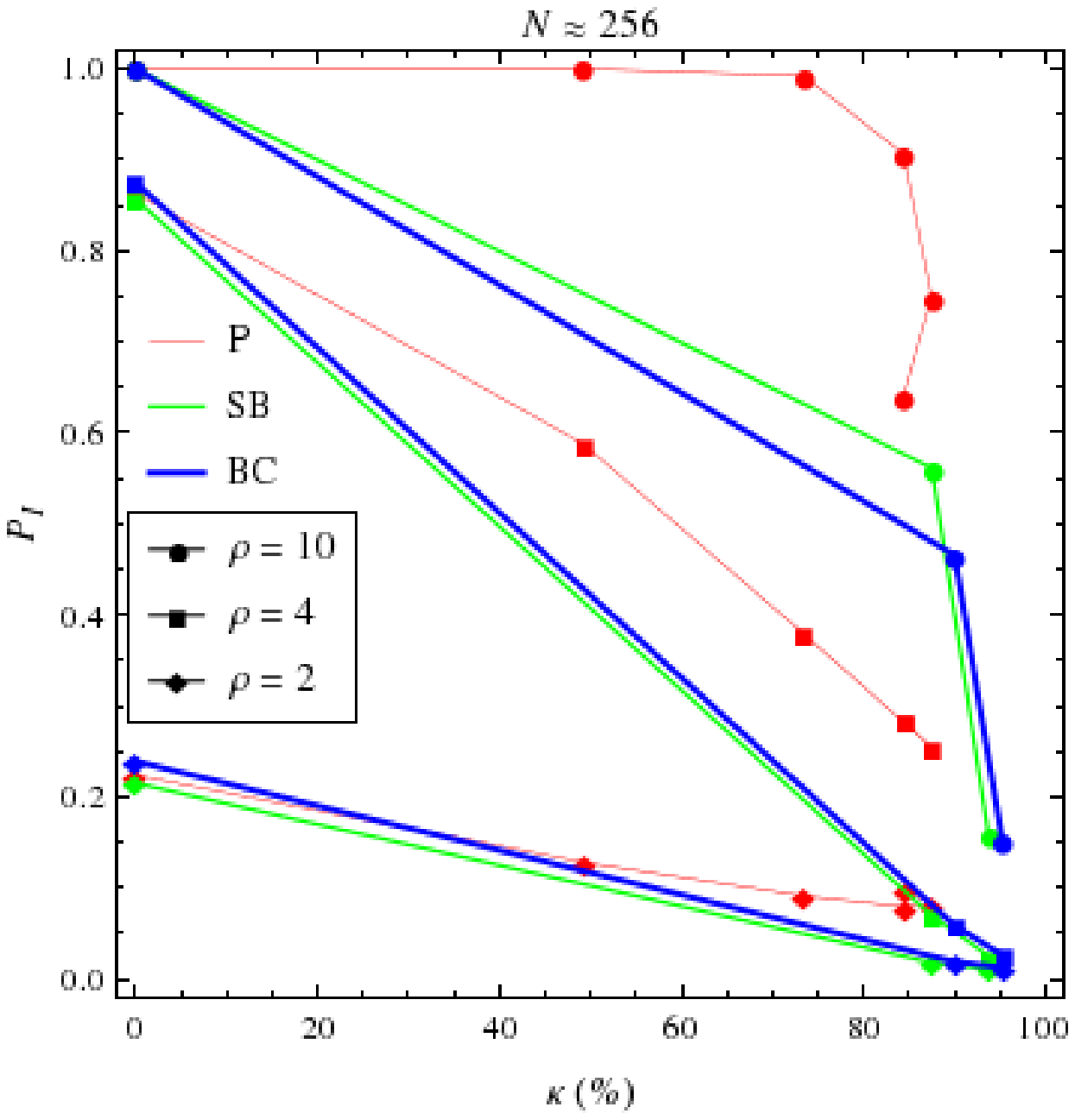}}
\caption{Plots of (a) discrimination $D$ and (b) accurate-identification rate $P_I$ against compression rate $\kappa$ for the partition, symmetric base and binomial coefficient schemes, at different values of true SNR $\rho$ for a $\approx256$-template bank.}
\label{fig:compareperformance}
\end{figure*}

In this section, we compare the performance of the uncorrelated partition scheme and its two correlated alternatives across three areas: template bank compression, GW detection and GW identification (i.e. localisation to a single template). The detection and identification plots here (and throughout the rest of the paper) were obtained using $10^5$-trial Monte Carlo simulations, and so the errors on each plot point are $\sim10^{-3}$ for a one-sigma binomial confidence interval.

Log--log plots of $M$ against $N$ for various conic compression schemes are shown in Fig.\,\ref{fig:comparecompression}, where the maximum lossless compression provided by the binary scheme \cite{W2014} has also been included for reference. As alluded to in Secs\,\ref{subsec:partition}--\ref{subsec:binomial}, the partition scheme has the largest range of compression rates, both in terms of compression bounds (plot area) and admitted rates (discrete density, not shown). The two correlated schemes cover similar areas at lower densities in Fig.\,\ref{fig:comparecompression}, with the binomial coefficient scheme offering slightly greater compression.

Detection performance for each compression setting of a given scheme may be measured by detection sensitivity at a fixed false alarm rate (which is simply read off the corresponding ROC curve), or by a summary statistic that captures most of the information contained in an ROC curve (e.g. the area $A_\mathrm{ROC}$ under the curve). Since an ROC curve always lies above the no-discrimination line $P_D=P_F$, we define the discrimination
\begin{equation}\label{eq:discrimination}
D:=2A_\mathrm{ROC}-1,
\end{equation}
which serves as a measure of how well the detection statistic discriminates between true and false positives.

Fig.\,\ref{fig:compareperformance}(a) shows plots of discrimination against compression for the three proposed schemes at different values of true SNR, with $N\approx256$. We use the maximum-overlap detection statistic in lieu of the optimal statistic for the partition scheme, and are compelled to choose $N=210$ for the binomial coefficient scheme. The three schemes have comparable performance at lower SNRs, but the partition scheme begins to outpace its correlated alternatives as SNR increases.

To compare identification performance (after a true detection), we consider plots of accurate-identification rate $P_I$ against compression, but only for the fastest standard algorithms of each scheme (i.e. $\mathrm{I}_1$ for the partition scheme, and automatic identification $\mathrm{I}_0$ for the correlated schemes). The rate $P_I$ for each plot point is calculated using all and only the trials with the injected signal present, and therefore assumes perfect detection throughout ($P_D=1$ and $P_F=0$). This decouples identification from detection: it allows standardised comparison of the schemes at a fixed false alarm rate, and does not penalise the identification performance of any method for having inferior detection performance.

As seen in Fig.\,\ref{fig:compareperformance}(b), the usefulness of lossless compression and automatic identification is limited in the presence of noise; the addition of a simple fine-grained search to the partition scheme is enough to yield significantly higher identification accuracy even at marginally lower compression. The turnaround in accurate-identification rates for the partition scheme at larger values of $P$ is due to the additional statistic evaluations used in the fine-grained search, which for $\mathrm{I}_1$ gives $N_\mathrm{eval}=M+P$ in \eqref{eq:compression}. Since $M=N/P$, $\kappa(P)$ has one turning point. For this example, $P=8$ and $P=64$ provide the same level of compression; identification accuracy is higher for the former at $\rho=10$, similar for both at $\rho=4$, and higher for the latter at $\rho=2$.

In summary, the partition scheme offers better overall performance than its correlated alternatives at the same level of compression. For GW detection, the introduced correlations among the conic statistics lead to slightly reduced detection sensitivity and discriminatory power at high SNR; furthermore, the potential benefits of lossless compression for GW identification turn out to be nullified by the effects of noise. Hence there appears to be little reason for using correlated schemes over the partition scheme, which is more promising as it is easy to implement and admits a relatively populated sliding scale of compression rates. We further investigate and implement the partition scheme as the representative conic compression scheme in Secs\,\ref{sec:assumptions} and \ref{sec:example}.

\section{Orthogonality and subspaces}\label{sec:assumptions}

The conic compression schemes proposed in Sec.\,\ref{sec:schemes} are fully general, in the sense that no prior assumptions about the template bank are made apart from \eqref{eq:orthogonality} and \eqref{eq:1Dsubspace}. These orthogonal and 1-D restrictions are neither realistic nor optimal, as template banks typically feature highly correlated neighbouring templates and are unlikely to contain a template exactly proportional to the GW signal itself. In this section, we discuss the (separate) lifting of each assumption for the partition scheme, and the resultant effects on detection sensitivity and localisation accuracy. Each approach may be viewed as a simplified limiting case of an actual template bank, which can always be made dense enough to include a signal-proportional template (assuming model accuracy), or orthogonalised. A more realistic example with both assumptions lifted is considered in Sec.\,\ref{sec:example}.

\subsection{Non-orthogonal templates}\label{subsec:nonorthogonality}

\begin{figure*}
\centering
\captionsetup[subfloat]{position=top}
\subfloat[]{\includegraphics[width=0.9\columnwidth]{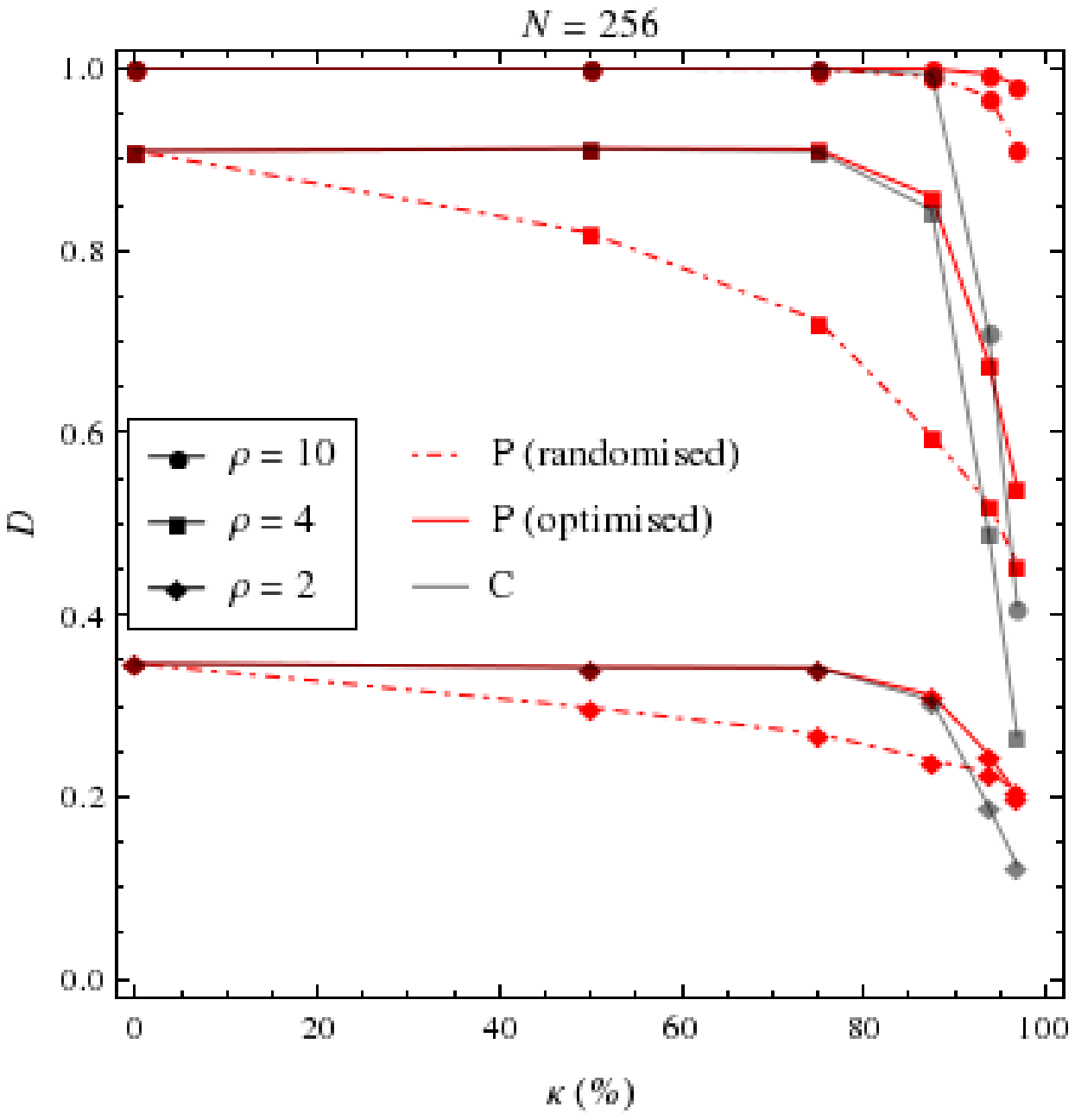}}
\subfloat[]{\includegraphics[width=0.9\columnwidth]{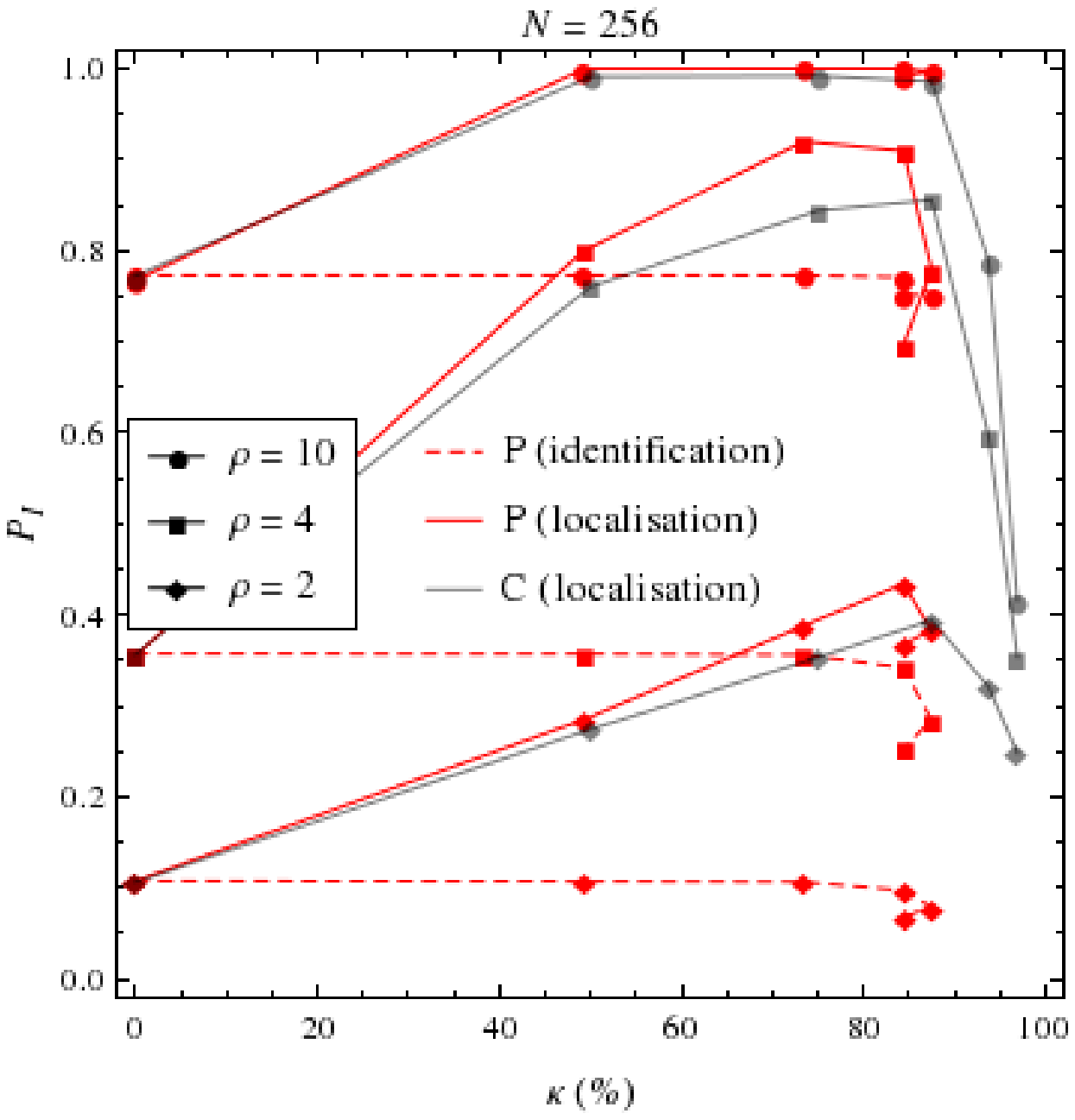}}
\caption{Plots of (a) discrimination $D$ against compression rate $\kappa$ for the randomised/optimised partition scheme and the coarsening method, and (b) accurate-identification/localisation rate $P_I$ against compression rate $\kappa$ for the optimised partition scheme and the coarsening method, at different values of true SNR $\rho$ for a non-orthogonal 256-template bank. Accurate localisation here is defined as the identification of the template $h_1$ or one of the nearest $P=1/(1-\kappa)$ templates.}
\label{fig:orthogonality}
\end{figure*}

We first consider a sufficiently dense bank of correlated (non-orthogonal) templates, such that the GW signal still lies in the 1-D subspace spanned by a single template in Hilbert space. From the first equalities in \eqref{eq:originalexpectation}, \eqref{eq:originalcovariance}, \eqref{eq:conicexpectation} and \eqref{eq:coniccovariance}, it follows in the presence of a GW signal that
\begin{equation}\label{eq:nonorthogonalexpectation}
\mathrm{E}(X_m)=A\sum_{n\in\mathbf{U}_m}\langle h_1|h_n\rangle,
\end{equation}
\begin{equation}
\mathrm{cov}(X_m,X_{m'})=\sum_{n\in\mathbf{U}_m}\sum_{n'\in\mathbf{U}_{m'}}\langle h_n|h_{n'}\rangle.
\end{equation}
Any partition of $\mathbf{N}$ as in Sec.\,\ref{subsec:partition} defines a splitting of the (sorted) original mean vector and covariance matrix into $P\times1$ blocks and $P\times P$ blocks respectively; each entry in the conic mean vector and covariance matrix is then simply the sum of entries in the corresponding block, which reflects the coarse-graining of the compression.

As a toy model for investigating non-orthogonal templates, we use a frequency-parametrised bank of sinusoidal waveforms $h=\sin{(2\pi ft)}$ with finite observation time $T$. Assuming white noise for simplicity, the inner product \eqref{eq:innerproduct} may be written as
\begin{equation}
\langle\mathcal{X}|\mathcal{F}\rangle\propto\int_0^T\mathcal{X}(t)\mathcal{F}(t)\,dt.
\end{equation}
For an $N$-template bank with $f_\mathrm{min}\leq f\leq f_\mathrm{max}$ and $\delta f:=(f_\mathrm{max}-f_\mathrm{min})/(N-1)\ll f_\mathrm{min}$, the overlaps are given by

\begin{widetext}
\begin{equation}\label{eq:overlap}
\langle h_n|h_{n+\Delta n}\rangle\approx2\int_0^1\sin{(2\pi f_\mathrm{min}t)}\sin{(2\pi(f_\mathrm{min}+|\Delta n|\delta f)t)}\,dt,
\end{equation}
\end{widetext}

\noindent where we have normalised to $T=1$ such that $f$ is given in waveform cycles per observation time. This sinc-like function of $\Delta n\in\mathbb{Z}$ yields a band covariance matrix for $x_n$; we set $N=256$, and choose the frequency bounds such that $\mathrm{cov}(x_n,x_{n\pm 1})\approx0.97$ \cite{DS1994,SD1991,DS1994a} (i.e. a maximal mismatch \cite{O1996} of around 0.03).

In contrast to the orthogonal case, the choice of partition generally affects the performance of the partition scheme for non-orthogonal templates. For the one-parameter template bank with overlaps given by \eqref{eq:overlap}, we consider both a randomised partition and a more optimised (but not necessarily optimal) partition with $\mathbf{U}_m=\{n\in\mathbb{Z}^+\,|\,(m-1)P<n\leq mP\}$. We also include for comparison a uniformly spaced $M$-template subset of the original bank (equivalently, $\mathbf{U}_m=\{n\in\mathbb{Z}^+\,|\,n=mP\}$ where $\bigcup_{m\in\mathbf{M}}\mathbf{U}_m\neq\mathbf{N}$). This ``coarsened'' template bank is not compressed; it is obtained in a more straightforward way by simply reducing the correlation (increasing the maximal mismatch) between neighbouring templates. The standard detection algorithm outlined in Sec.\,\ref{subsec:partition} is then applied for the two partition schemes and the coarsening method.

Fig.\,\ref{fig:orthogonality}(a) shows plots of discrimination (using $X_\mathrm{max}$) against compression for both choices of partition and the coarsened template bank, where performance in the presence of a GW signal is averaged over the $N$ possible locations of the corresponding template in the bank. The optimised partition (with highly correlated templates grouped together) outperforms its randomised counterpart at all considered values of true SNR. It also shows significant improvement over the coarsening method at higher compression rates, which is expected as it uses information from the full $N$-template bank rather than just an $M$-template subset.

The largest statistic evaluation for the coarsened template bank identifies a best guess for the GW signal, but the accuracy of this identification is zero if the signal does not correspond to a template in the coarsened bank. Since the spacing of the coarsened bank is $P$, we may consider the best-guess template as representative of the $P$ templates nearest to it (or $P-1$ if $P$ is odd), and say that the largest statistic evaluation localises a best guess for the signal. We then define the localisation to be accurate if the correct template $h_1$ is one of those templates (equivalently, if the identified best-guess template is $h_1$ or one of the $P$ templates nearest to $h_1$). The identification algorithms in Sec.\,\ref{subsec:partition} also identify a single best-guess template for the partition scheme, which allows us to consider both accurate identification (to a precision of 1) and accurate localisation (to a precision of $P$) in the same way. Fig.\,\ref{fig:orthogonality}(b) shows plots of accurate-identification and localisation rates (using $\mathrm{I}_1$, which gives $N_\mathrm{eval}=M+P$ in \eqref{eq:compression}) against compression for the (optimised) partition scheme and the coarsening method.

As in Sec.\,\ref{subsec:comparison}, the turnaround in accurate-identification and localisation rates for the partition scheme is due to the additional statistic evaluations of the fine-grained search. The localisation rates increase up to some level of compression, which is mainly because ``accurate'' localisation is defined up to a degree of precision that degrades with compression; this effect is seen for the coarsening method as well. Localisation to within the spacing of the original template bank (i.e. identification) decreases monotonically in accuracy for the partition scheme, and will not be achievable for the majority of signals with the coarsening method. The partition scheme localises the GW signal with slightly greater accuracy than the coarsening method, and in fact identifies it with virtually no fall-off in accuracy at significant compression levels.

Increasing the correlation between neighbouring templates is known to improve the detection and localisation performance of a general template bank \cite{DS1994,SD1991,O1996}. Results in this section illustrate that the partition scheme retains these benefits up to high levels of compression, and provides a superior alternative to simply coarsening the template bank for computational savings. The viability of conic compression becomes even more evident in Sec.\,\ref{sec:example}, where we apply the partition scheme to a larger and more broadly correlated two-parameter template bank.

\subsection{2-D subspace}\label{subsec:2Dsubspace}

\begin{figure*}
\centering
\captionsetup[subfloat]{position=top}
\subfloat[]{\includegraphics[width=0.9\columnwidth]{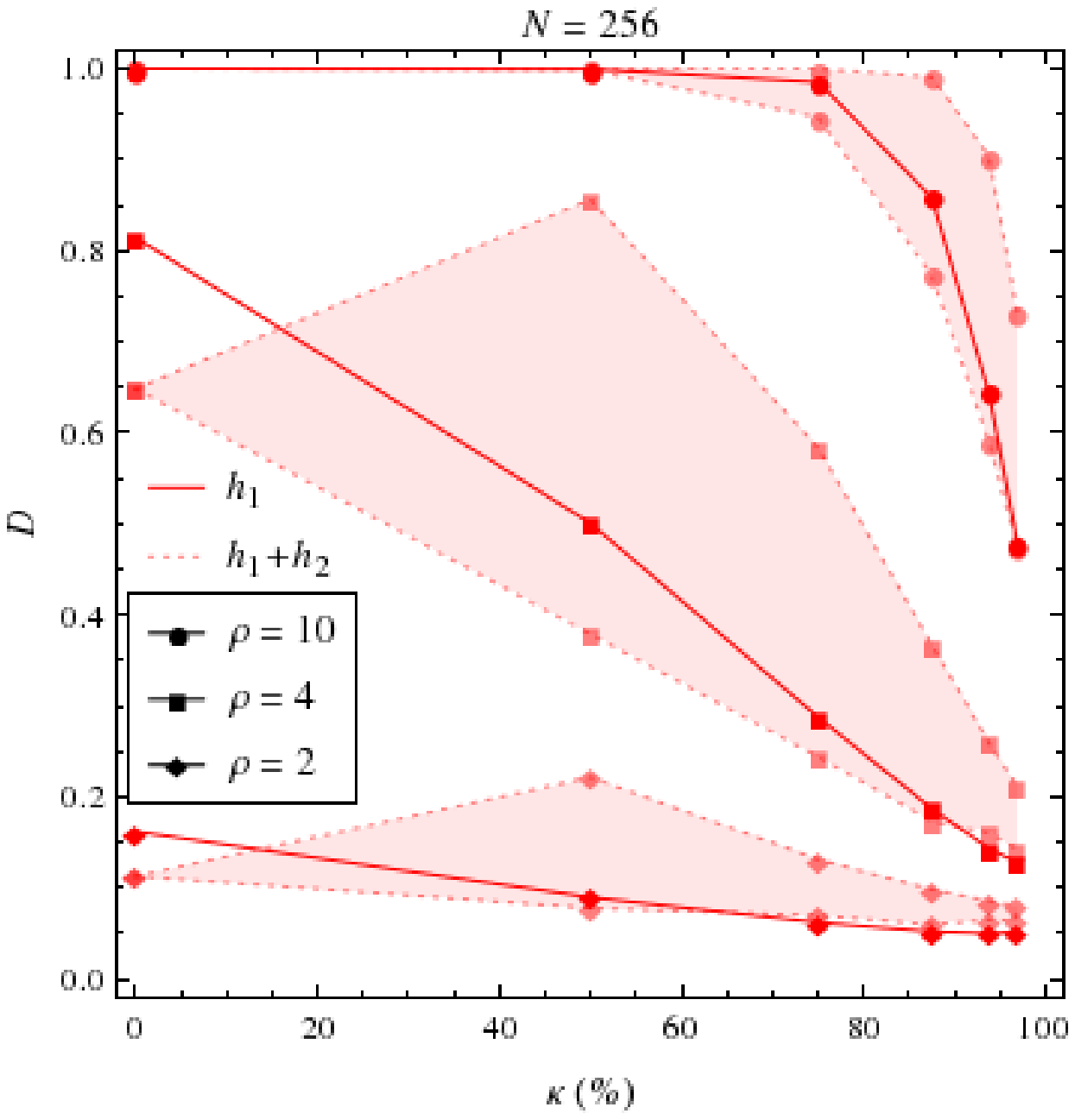}}
\subfloat[]{\includegraphics[width=0.9\columnwidth]{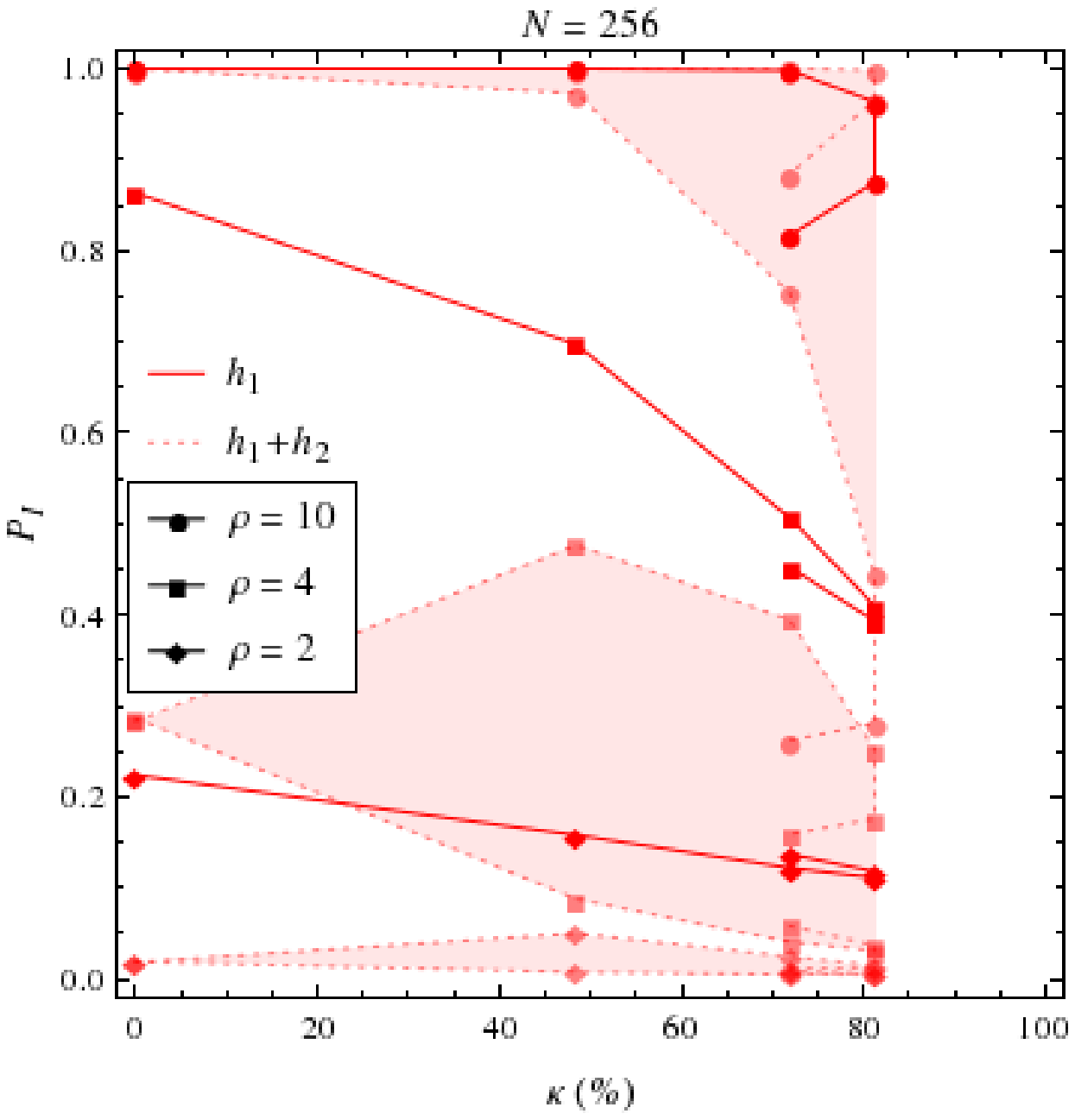}}
\caption{Plots of (a) discrimination $D$ and (b) accurate-identification rate $P_I$ against compression rate $\kappa$ for the partition scheme with 1-D and 2-D signals, at different values of true SNR $\rho$ for a 256-template bank. The higher dotted curves for each value of $\rho$ correspond to the template labels 1 and 2 being assigned to the same set, while the lower curves correspond to them being assigned to different sets. Accurate identification for the 2-D case is defined as the identification of both templates $h_1$ and $h_2$.}
\label{fig:subspaces}
\end{figure*}

Throughout Secs\,\ref{sec:schemes} and \ref{subsec:nonorthogonality}, we have assumed that the GW signal is exactly proportional to a template in the bank. To understand the impact on compression performance when this is not the case, we consider a bank of $N$ uncorrelated templates obtained through some orthogonalisation procedure (e.g. as in \cite{H2009,CEA2010,FEA2011}) on a general template bank, and a signal lying in the $N$-dimensional Hilbert space spanned by the orthogonal set. If $N$ is large, the signal is typically restricted to a low-dimensional subspace (this follows from the volume of an $N$-sphere). For simplicity, we assume it lies exactly between two templates in a 2-D subspace, i.e.
\begin{equation}\label{eq:2Dsubspace}
\mathcal{S}(t)=A(h_1(t)+h_2(t)),
\end{equation}
where the templates have been relabelled without loss of generality and $A=\rho/\sqrt{2}$ from \eqref{eq:SNR}. Hence the expectation values of the original and conic statistics become
\begin{equation}
\mathrm{E}(x_n)=A(\delta_{1n}+\delta_{2n}),
\end{equation}
\begin{equation}
\mathrm{E}(X_m)=A\,\mathrm{card}(\{1,2\}\cap\mathbf{U}_m),
\end{equation}
while their covariances remain as \eqref{eq:originalcovariance} and \eqref{eq:coniccovariance} respectively. The assumption \eqref{eq:2Dsubspace} is the worst-case scenario for a 2-D subspace, since the signal is maximally far from both templates in the subspace.

Although it is not possible to pre-optimise the choice of partition for orthogonal templates, the performance of the partition scheme in the presence of a 2-D GW signal \eqref{eq:2Dsubspace} falls into two partition-dependent cases. At small values of $P$, it is more likely that the labels $1\in\mathbf{U}_m$ and $2\in\mathbf{U}_{m'}$ are assigned to different sets ($m\neq m'$); as $P$ increases, so does the probability that they are assigned to the same set ($m=m'$), which improves performance (e.g. the effective SNR for detection purposes is raised by a factor of $\sqrt{2}$).

The standard detection algorithm in Sec.\,\ref{subsec:partition} is applicable for a 2-D signal, while the standard identification algorithms may be generalised at step (iv) by considering the two largest original statistic evaluations instead. Fig.\,\ref{fig:subspaces} shows plots of discrimination and accurate-identification rate against compression for a 2-D signal $\mathcal{S}\propto h_1+h_2$, compared against a 1-D signal $\mathcal{S}\propto h_1$ with the same true SNR $\rho$. The identification algorithm $\mathrm{I}_2$ is used, since the accuracy rate of $\mathrm{I}_1$ falls to zero if $m\neq m'$. This gives $N_\mathrm{eval}=M+2P$ in \eqref{eq:compression}.

For detection of a 2-D GW signal, the effectiveness of the partition scheme is reduced slightly at lower SNRs, but mitigated by the case where $m=m'$ (i.e. the higher dotted curves in Fig.\,\ref{fig:subspaces}). Detection performance for this special case actually improves up to some level of compression, which is possible as the symmetry among all possible signals is broken (by the partitioning process). A similar effect is seen for the example in Sec.\,\ref{sec:example}. The discrimination for a 1-D signal generally lies within the 2-D discrimination bounds; at higher compression rates, there is little to no detection performance lost if the signal is not confined to a 1-D subspace.

Accurate identification of a 2-D GW signal (i.e. the identification of both $h_1$ and $h_2$, in this toy model) is more problematic than in the 1-D case, since accuracy rates are reduced to begin with and fall off rapidly even at high SNR. Nevertheless, options such as lowering compression or switching to $\mathrm{I}_{i>2}$ are available for the partition scheme, which should at least allow the template with maximal signal overlap to be identified at acceptable accuracy rates.

If the true SNR is sufficiently high, the standard algorithms $\mathrm{I}_{i>j}$ may also be used to identify a $j$-dimensional GW signal described by an arbitrary linear combination of templates, i.e.
\begin{equation}
\mathcal{S}(t)=\sum_{k=1}^jA_kh_k(t),
\end{equation}
where $A_k>A_{k+1}$ and the templates have been relabelled without loss of generality. At step (iv) of the algorithms, each ordered weight $A_k$ may be approximated by the $k$-th largest original statistic evaluation $x_{(k)}$, with the SNR of the identified signal given by
\begin{equation}
\rho_I=\sqrt{\sum_{k=1}^jx_{(k)}^2}.
\end{equation}
While this method fully recovers the (relative) weights of a GW signal's $j$ largest modes in the limit of infinite true SNR, its accuracy might be limited for lower-SNR signals and/or large values of $P$.

\section{Example: Taylor-T2 template bank}\label{sec:example}

\begin{figure}
\centering
\includegraphics[width=0.9\columnwidth]{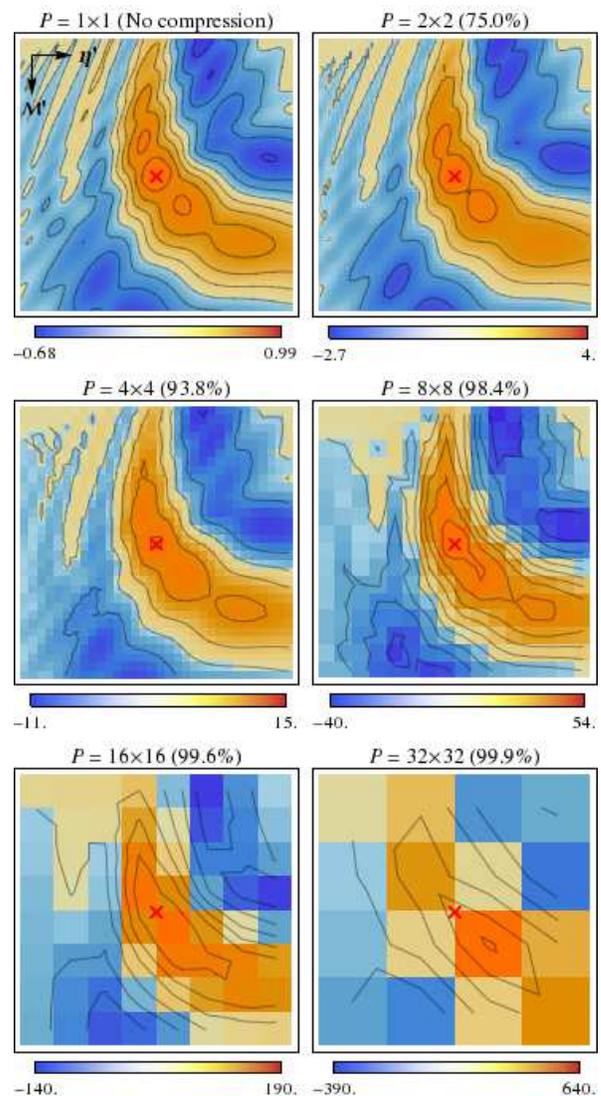}
\caption{Matrix/contour plots of the expectation values $\mathrm{E}(x_n)$ and $\mathrm{E}(X_m)$ for the partition scheme, at different values of set cardinality $P$ (with compression rate $\kappa$ in parentheses) for a Taylor-T2 GW signal (red cross) injected between the four central templates of a ($128\times128$)-template Taylor-T2 bank. The signal has chirp mass $\mathcal{M}=10^6\,M_\odot$ and symmetric mass ratio $\eta=0.15$, while the bank is gridded uniformly in linearly transformed parameters $\mathcal{M}'$ (increasing from top to bottom) and $\eta'$ (increasing from left to right) with maximal mismatch $\approx0.01$. Overlap values depend on the true SNR $\rho$ (set to 1 in these plots), and range from positive (orange) to negative (blue) in some subinterval of $(-P\rho,P\rho)$.}
\label{fig:overlaps}
\end{figure}

\begin{figure*}
\centering
\captionsetup[subfloat]{position=top}
\subfloat[]{\includegraphics[width=0.9\columnwidth]{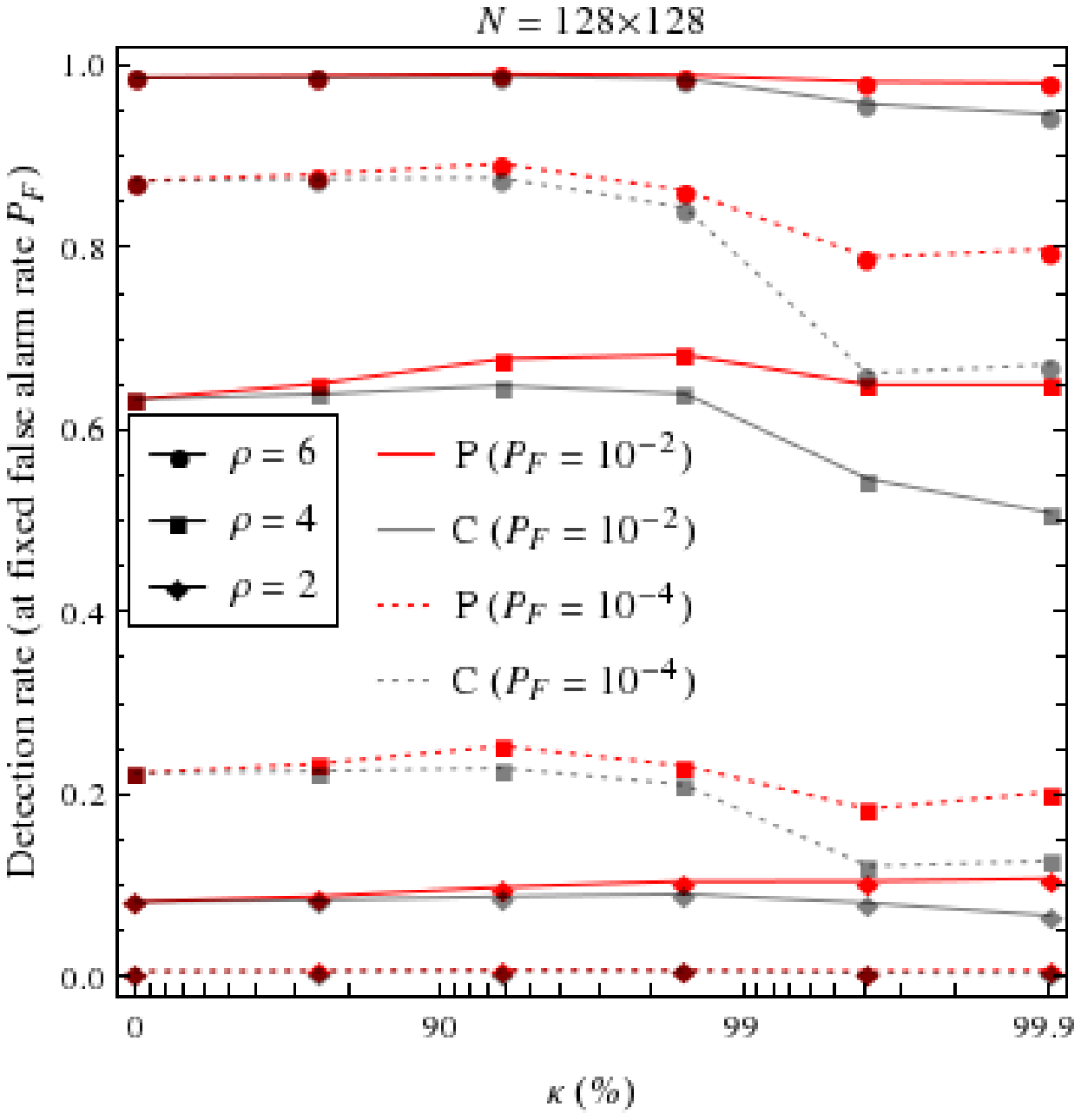}}
\subfloat[]{\includegraphics[width=0.9\columnwidth]{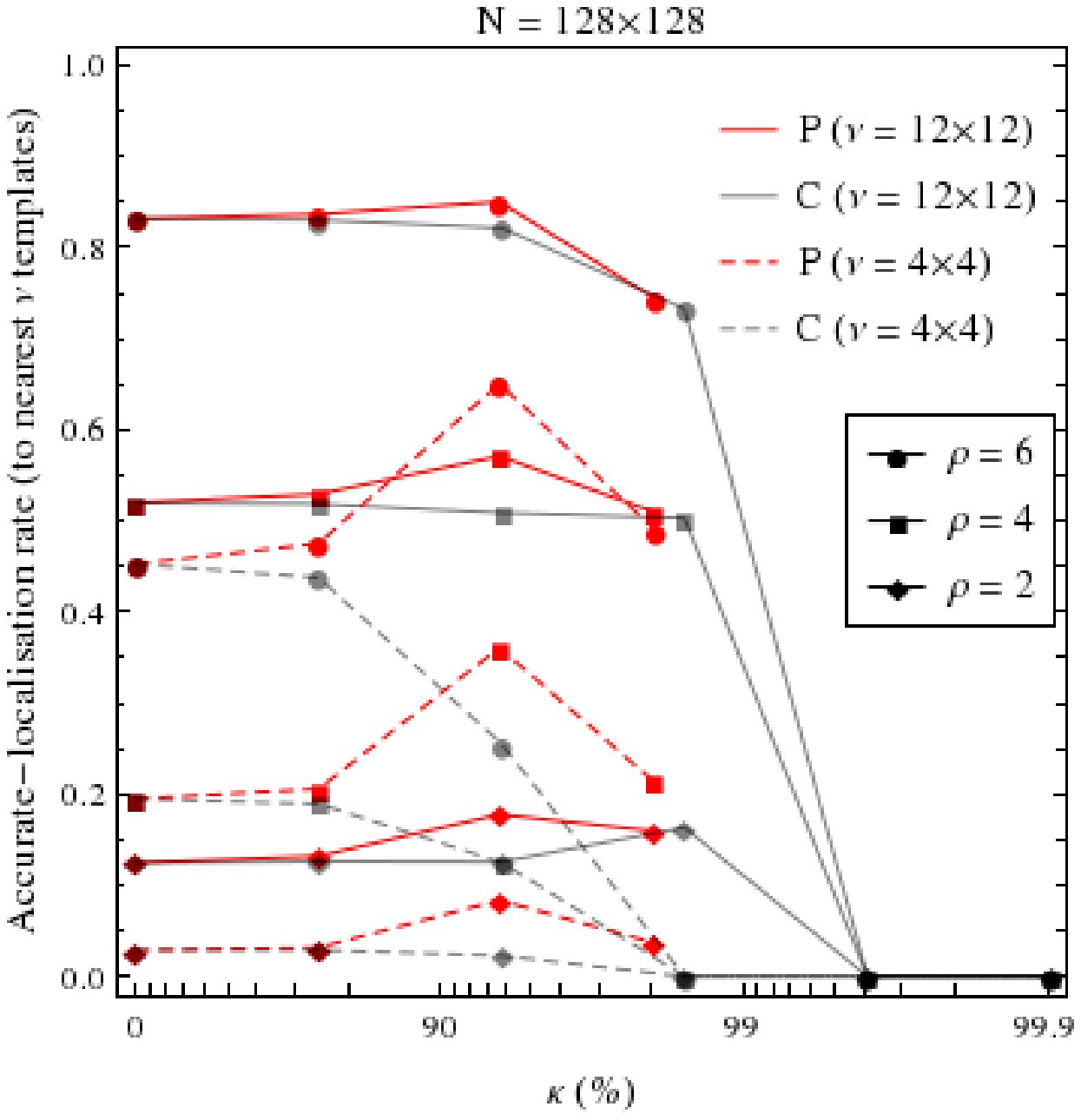}}
\caption{Plots of (a) detection rate $P_D$ (at fixed false alarm rate $P_F$) and (b) accurate-localisation rate $P_I$ (to nearest $\nu$ templates) against compression rate $\kappa$ for the optimised partition scheme and the coarsening method, at different values of true SNR $\rho$ for a GW signal injected with central parameters $\boldsymbol{\theta}_C$. Accurate localisation here is defined as the identification of a template within the central squares of $\nu$ templates.}
\label{fig:taylort2}
\end{figure*}

In this section, we implement the (optimised) partition scheme described in Secs\,\ref{subsec:partition} and \ref{subsec:nonorthogonality} for a larger and more realistic example: a two-parameter template bank of mixed-order PN waveforms, which describe the gravitational radiation emitted during the inspiral part of a comparable-mass binary merger. An optimised partition in this case (and in general) refers to a partition of the template bank such that highly correlated templates are grouped together as much as possible.

The waveform family we use is the Taylor-T2 approximant \cite{BIWW1996,B2014} for a circular and non-inclined binary with comparable component masses $m_1\geq m_2$. These waveforms are parametrised by their chirp mass $\mathcal{M}=(m_1m_2)^{3/5}/(m_1+m_2)^{1/5}$ and symmetric mass ratio $\eta=m_1m_2/(m_1+m_2)^2$, and are written as PN expansions in the frequency-related variable $x=(G\mathcal{M}\eta^{-3/5}\dot{\phi}/c^3)^{2/3}$, where $\dot{\phi}$ is the time derivative of the orbital phase $\phi$. We truncate the PN expansions at finite order, specifying the phase, amplitude and mass monopole to 3.5PN, 2PN and 1PN respectively; the resultant mixed-order waveform may be written compactly as \cite{CG2014}
\begin{equation}\label{eq:taylort2}
h_{\mathcal{M},\eta}(t)=\frac{2G\mathcal{M}\eta^{2/5}}{c^2R}A(t)e^{2i\psi(t)},
\end{equation}
where $R$ is the source distance (which the true SNR $\rho$ is inversely proportional to), and expressions for the amplitude function $A$ and tail-distorted orbital phase $\psi$ are given in App. \ref{app:PN}.

Template bank compression is potentially more important for analysing data from the low-frequency eLISA detector, since the long duration of sources in the eLISA band results in a much larger number of templates required to cover parameter space \cite{GEA2004}. As mergers of massive black-hole binaries are an anticipated source for eLISA \cite{AEA2012}, we consider a Taylor-T2 GW signal with the parameters $\boldsymbol{\theta}_C=(1,0.15)$, where $\boldsymbol{\theta}:=(\mathcal{M}/(10^6\,M_\odot),\eta)$; this corresponds to a binary black-hole inspiral with component masses $(m_1,m_2)=(1.9\mathcal{M},0.7\mathcal{M})$. The duration of the signal is set to $t_c=1\,\mathrm{yr}$.

We also generate a bank of Taylor-T2 templates with the same duration, each normalised with respect to the inner product \eqref{eq:innerproduct}, where $S_\mathcal{N}(f)$ is given by a (two-sided) analytic approximation to the eLISA noise power spectral density \cite{AEA2013}. These templates are gridded uniformly in the transformed parameters $\boldsymbol{\theta}':=\boldsymbol{\theta}_C+L(\boldsymbol{\theta}-\boldsymbol{\theta}_C)$ (128 points in each parameter), with the signal lying in the middle of the four central templates and the linear transformation $L$ chosen such that the template overlaps are isotropic with respect to the grid (at least for the central region). The maximal mismatch of each template with its four nearest neighbours is around 0.01.

Since the $N=16384$ templates are pre-sorted by the (skewed) square grid, an optimised (but not necessarily optimal) partition is obtained by the obvious grouping into $M$ blocks of $\sqrt{P}\times\sqrt{P}$ templates. This particular template bank admits six nontrivial square partitions with $P\in\{4,16,64,256,1024,4096\}$; we do not consider the case $P=4096$, as $P=1024$ already yields a compression rate of $99.9\%$. A large number of rectangular partitions (where $P=2^i$ with $0<i<14$) are also possible, but we omit these here for simplicity as they are degenerate with the square partitions and among themselves. Square partitions are straightforward to generalise for various lattice choices \cite{BCCS1998,OS1999,P2007}, and will be fairly optimal as long as the templates are gridded uniformly in the parameter-space metric.

The expectation values of the original and conic statistics (the first equality in \eqref{eq:originalexpectation} and \eqref{eq:nonorthogonalexpectation} respectively) are visualised in Fig.\,\ref{fig:overlaps}, where the coarse-graining of the compression is evident. Overlaps for the Taylor-T2 template bank are much less localised than the toy model overlaps in Sec.\,\ref{subsec:nonorthogonality}; this is due to their wider cycle widths in both $\mathcal{M}$ and $\eta$, as well as a slight degeneracy in the two parameters (overlaps at the boundary of the first plot in Fig.\,\ref{fig:overlaps} can be as high as $0.4\rho$). As the templates are so broadly correlated and the GW signal is injected right in the centre of the bank, the partition scheme is expected to perform well up to a high level of compression.

For comparison purposes, we again consider the simple coarsening method discussed in Sec.\,\ref{subsec:nonorthogonality}. The smaller coarsened banks are formed by selecting individual templates near the centre of each square block in the original bank, rather than by summing the templates in each block (as in the partition scheme). Detection and localisation performance for both the partition scheme and the coarsening method on the Taylor-T2 template bank with central injection is summarised in Fig.\,\ref{fig:taylort2}. The semi-log plots in this section use an abscissa of $-\lg{(1-\kappa)}/3$, as most of the considered compression rates are $>90\%$.

Instead of the discrimination \eqref{eq:discrimination}, we quantify detection performance using the detection rates at two fixed false alarm rates $P_F=10^{-2}$ and $P_F=10^{-4}$ (the number of Monte Carlo trials performed for each plot point is $\sim10^5$, and so the errors are $\sim10^{-3}$ for a one-sigma binomial confidence interval). At all considered values of SNR and fixed false alarm rate, there is no fall-off in the partition scheme's detection performance up to $\kappa=93.8\%$ (and even a slight increase, due to the special choice of central injection). While this is also the case for the coarsening method, detection rates for the partition scheme are distinctly higher at compression rates of $>90\%$, with improvements of over 0.1 at $\kappa=99.9\%$.

The identification algorithm $\mathrm{I}_1$ is used to localise the GW signal, which gives $N_\mathrm{eval}=M+P$ in \eqref{eq:compression}. Rates for accurate localisation to within two central squares of $12\times12$ templates (corresponding to $<1\%$ of the entire bank) and $4\times4$ templates ($<0.1\%$ of the bank) are considered. Localisation is typically improved by compression up to $\kappa=93.7\%$, which is provided by two different values of $P$ (see discussion in Sec.\,\ref{subsec:comparison}). The two values are $P=16$, beyond which the matrix/contour plot of $\mathrm{E}(X_m)$ in Fig.\,\ref{fig:overlaps} loses scale-similarity to that of $\mathrm{E}(x_n)$, and $P=1024$, for which performance is regained as each conic template incorporates more of the original templates and accuracy is added by the fine-grained search. Localisation is poorer at $\kappa=98.0\%$, which corresponds to both $P=64$ and $P=256$. To reduce clutter in Fig.\,\ref{fig:taylort2}(b), only the higher localisation rates for $\kappa=93.7\%$ and $\kappa=98.0\%$ are plotted. The partition scheme outperforms the coarsening method at most levels of compression, especially in the case of accurate localisation to within the smaller square of $4\times4$ templates.

\begin{figure}
\centering
\includegraphics[width=0.9\columnwidth]{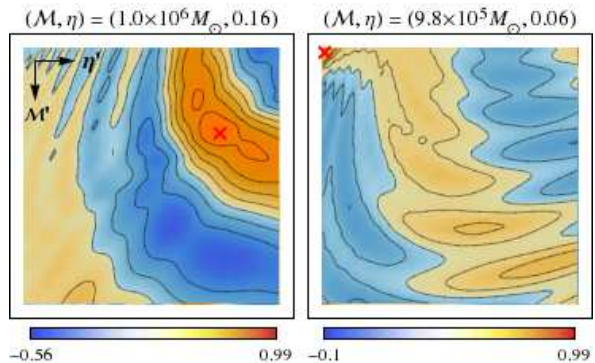}
\caption{Matrix/contour plots of the expectation values $\mathrm{E}(x_n)$ for Taylor-T2 GW signals (red crosses) injected in a ($128\times128$)-template Taylor-T2 bank at random (left) and near the boundary (right).}
\label{fig:injections}
\end{figure}

\begin{figure*}
\centering
\captionsetup[subfloat]{position=top}
\subfloat[]{\includegraphics[width=0.9\columnwidth]{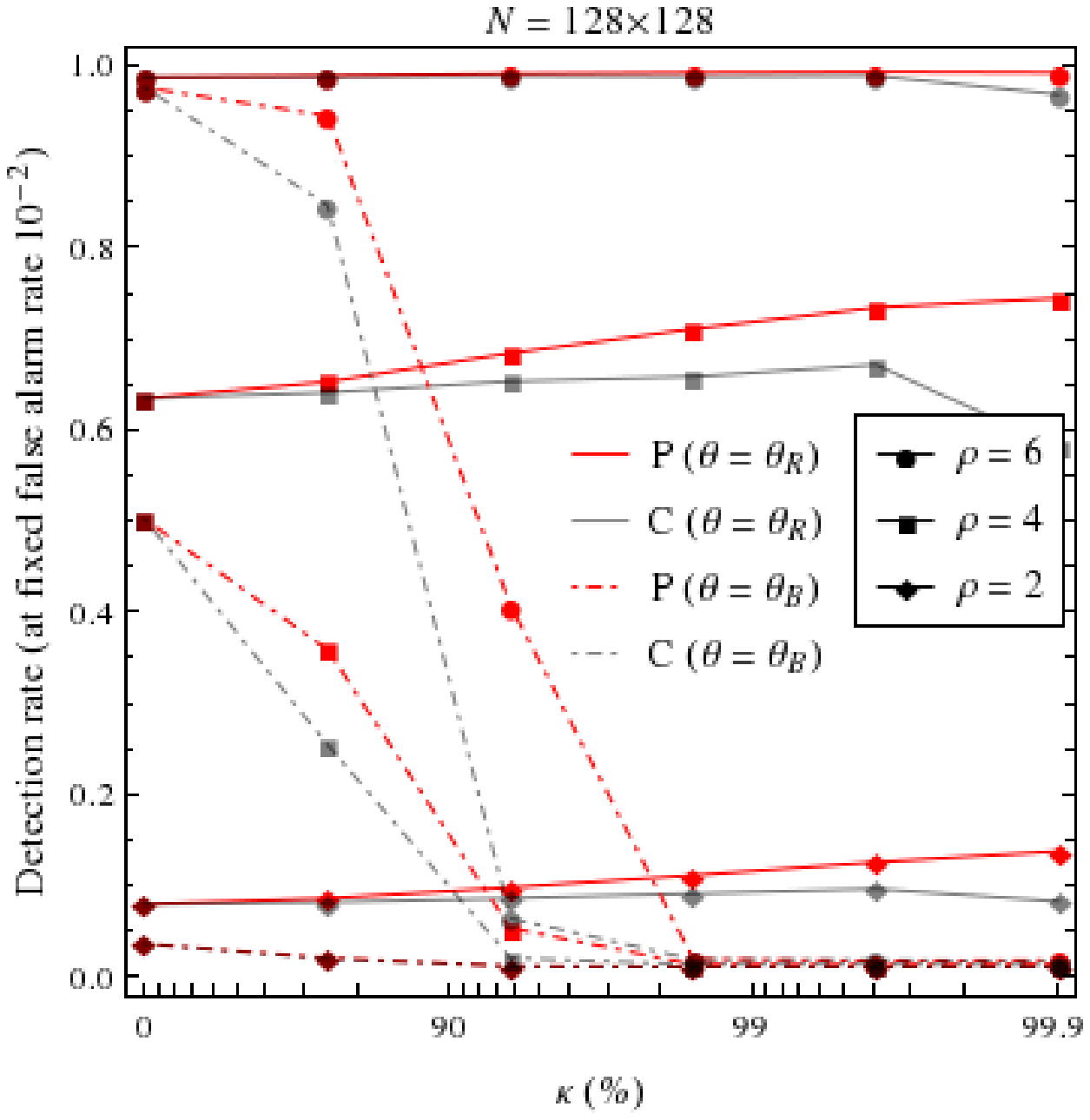}}
\subfloat[]{\includegraphics[width=0.9\columnwidth]{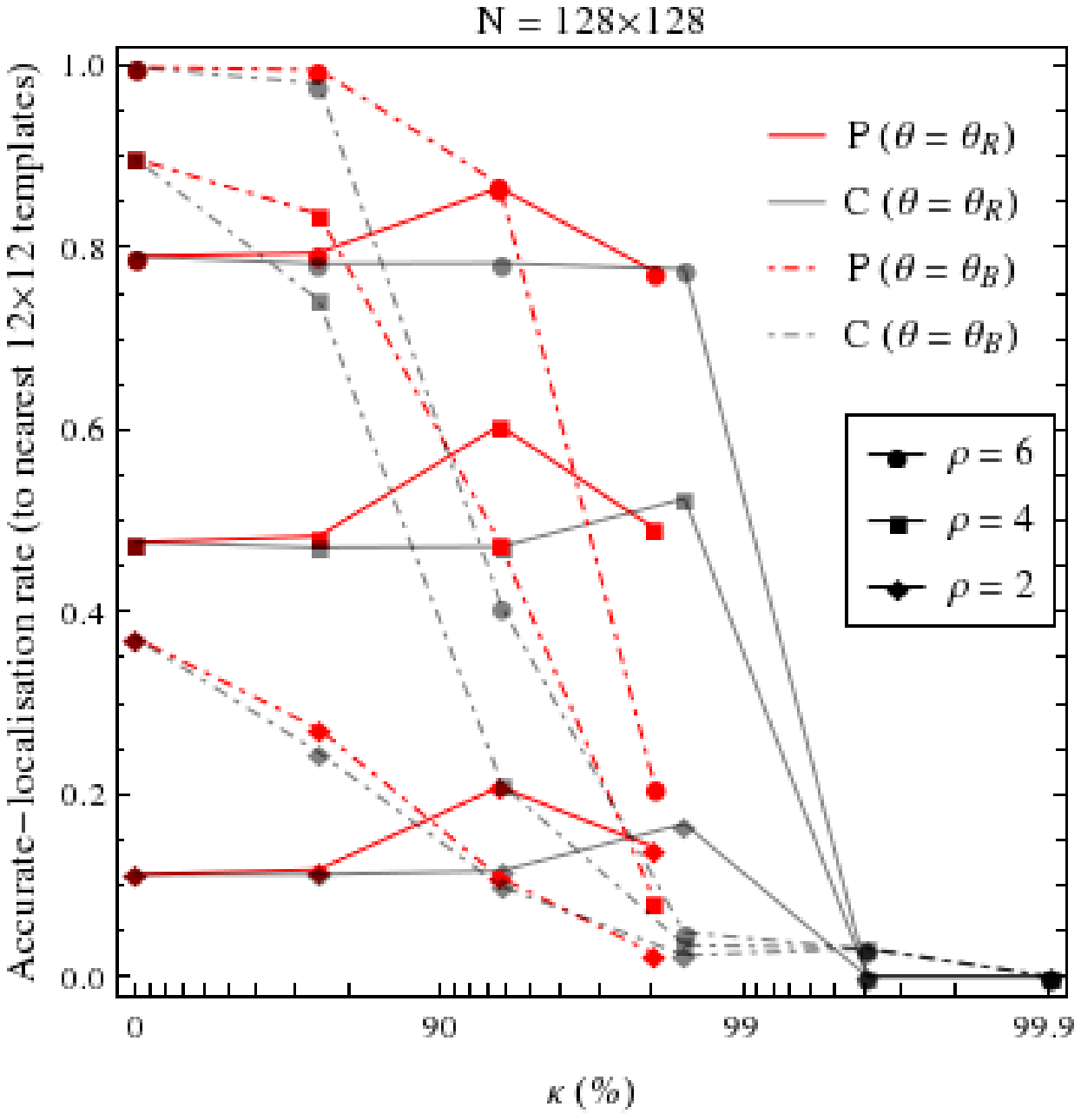}}
\caption{Plots of (a) detection rate $P_D$ (at fixed false alarm rate $P_F=10^{-2}$) and (b) accurate-localisation rate $P_I$ (to nearest $12\times12$ templates) against compression rate $\kappa$ for the optimised partition scheme and the coarsening method, at different values of true SNR $\rho$ for a GW signal injected with random parameters $\boldsymbol{\theta}_R$ and boundary parameters $\boldsymbol{\theta}_B$. Accurate localisation here is defined as the identification of a template within the square of $12\times12$ templates nearest to each injection.}
\label{fig:boundary}
\end{figure*}

For the special case of a centrally injected GW signal, the detection and localisation performance of the partition scheme is non-decreasing up to high levels of compression and can even rise above that of the original template bank; however, this may also be said for the coarsening method. To illustrate that the improvement of the partition scheme over the coarsening method is not simply due to the special choice of injection, we consider two other cases: a Taylor-T2 signal injected with randomly drawn parameters $\boldsymbol{\theta}_R=(1.0,0.16)$, and another injected near the boundary of the bank with the parameters $\boldsymbol{\theta}_B=(0.98,0.06)$ (i.e. in the middle of the four corner templates with low chirp mass and symmetric mass ratio). The expectation values of the original statistics for these two injections are visualised in Fig.\,\ref{fig:injections}.

Fig.\,\ref{fig:boundary} shows detection and accurate-localisation rates for both the partition scheme and the coarsening method on the random and boundary injections. The random injection is actually recovered with slightly better rates than the central injection rates in Fig.\,\ref{fig:taylort2}, but with a similar improvement of the partition scheme over the coarsening method. A more marked difference between the two methods is obtained for the boundary injection. Detection rates for both methods are now non-increasing, with the partition scheme showing greater improvement over the coarsening method; for the $\rho=6$ case, the improvement is around 0.3 at $\kappa=93.8\%$. Rates for accurate localisation of the boundary injection to within the corner square of $12\times12$ templates follow a similar trend, with a largest improvement of around 0.5 (again for the $\rho=6$ case at $\kappa=93.7\%$).

Detection and localisation performance for this Taylor-T2 example is injection-dependent, as it is for any realistic template bank: there is clearly no symmetry among all possible GW signals, since the templates are asymmetrically correlated and the signals may lie between templates. We have not undertaken a full injection-averaged analysis (similar to that performed in Sec.\,\ref{subsec:nonorthogonality}) due to the size of the template bank, but overall detection rates for such an analysis should decrease monotonically with compression as per intuition, with the partition scheme outperforming the coarsening method (as it does for the three injections presented here, as well as several others we have examined).

The partition scheme is expected to remain robust for searches in a ($d>2$)-dimensional parameter space. As the number of templates that are highly correlated with the GW signal increases exponentially with $d$, enlarging the span $P$ of each conic template at the same rate should maintain detection and localisation performance while increasing the relative computational savings (which scale as $1-1/P$). Good scaling with parameter-space dimensionality allows conic compression to be competitive with other search techniques that reduce computational cost. For example, the method of searching over time offset (i.e. signal time-of-arrival) using fast Fourier transforms yields a logarithmic reduction in the number of search points for that parameter \cite{BCCS1998}, but for multidimensional searches an overall logarithmic reduction is easily attained by the partition scheme with little impact on performance. The two methods might even be combined for greater savings, by constructing conic sums of templates aligned at a fixed reference time and using Fourier transforms of the conic templates to search over time offset.

\section{Conclusion}

In this paper, we have presented and compared three tunable conic compression schemes (partition, symmetric base and binomial coefficient) for a general template bank in a grid-based GW search. The bank is compressed in the preparatory offline stage, which yields faster detection and localisation of signals by reducing the number of inner product evaluations performed online.

A recently proposed binary labelling method \cite{W2014}, modified to ensure the equal treatment of templates, is contained as a particular case of the symmetric base scheme. Optimal detection statistics have been calculated for all three schemes under simplified conditions, and the standard maximum-overlap detection statistic (i.e. the maximum overlap over all the compressed templates) is shown to be significantly suboptimal for the base and binomial schemes. While these two lossless schemes provide automatic identification of the GW signal upon detection, the benefits of this are negated in the presence of noise; furthermore, the lossy partition scheme offers better detection and identification performance than its counterparts at the same level of compression.

We have applied the partition scheme to toy models of (i) a correlated template bank with a signal-proportional template and (ii) a signal lying in the span of orthogonal templates, to show that it remains feasible under such conditions. These toy models are instructive as they represent the two limiting cases of a general template bank. Correlations among the original templates result in partition-dependent performance, but this may be optimised beforehand by grouping highly correlated templates together; the optimised partition scheme is then superior to a simple coarsening of the template bank. If the signal is proportional to a linear combination of templates in an orthogonal bank, the detection performance of the scheme is not significantly reduced.

Conic compression performs well if the original template bank is sufficiently correlated, as demonstrated by our example implementation of the optimised partition scheme for a bank of $\sim10^4$ PN waveforms. We consider a centrally injected GW signal, a randomly injected one, and one at the boundary of the bank; again, the scheme is superior to the coarsening method across the board. The partition scheme is shown to be viable for practical applications, as it maintains good detection sensitivity and localisation accuracy up to high levels of compression and at all considered values of SNR for this more realistic template bank.

In summary, our tunable conic compression schemes---specifically the optimised partition scheme---might provide an effective method of improving the speed, detection sensitivity and localisation accuracy of GW template banks. The schemes are potentially useful for any search involving template banks, as they are fully general and may easily be adapted to supplement existing algorithms in GW data analysis pipelines. Conic compression is also particularly promising in the context of eLISA data analysis, where online grid searches are difficult as computational costs are more prohibitive; for example, the method could be used as an online tool to rapidly identify nearby sources before merger and generate alerts for electromagnetic telescopes.

\begin{acknowledgments}
We thank the anonymous referee for their constructive criticism, which has led to the improvement of this paper. AJKC's work was supported by the Cambridge Commonwealth, European and International Trust. JRG's work was supported by the Royal Society.
\end{acknowledgments}

\appendix

\section{Combinatorial design theory}\label{app:combinatorics}

The problem of constructing a family of sets $\mathbf{U}_m$ under the cardinality constraints \eqref{eq:constraintC} and \eqref{eq:constraintI} in Sec.\,\ref{subsec:binomial} may be regarded geometrically as the problem of constructing a collection of $N$ distinct points (representing template labels) and $M$ distinct lines (representing sets) with the following properties:
\begin{enumerate}[label=(\roman*)]
\item each point lies on exactly $R$ lines;
\item each line passes through exactly $C$ points;
\item any two lines intersect at exactly $I$ points;
\item any two points lie on at most $R-1$ lines.
\end{enumerate}
The final property is the automatic identification condition, i.e. no two labels are assigned to exactly the same subfamily of sets.

The feasibility of carrying out such a construction (or finding additional conditions on $N$, $M$ and $R$ that ensure it is possible) is a difficult and unsolved problem in combinatorics. One special case that has been studied in detail is $R=C$ and $I=1$. This implies that $N=M=R^2-R+1$, and that any two points must lie on exactly one line. Under these circumstances, the four geometrical properties define a finite projective plane of order $R-1$ \cite{B2002}. It is known that finite projective planes exist with prime orders \cite{B2002}, but there is no finite projective plane of order $6$ \cite{B1938} or $10$ \cite{LTS1989}, while the existence (or otherwise) of an order-$12$ finite projective plane remains an open question.

The special case of finite projective planes is uninteresting from a compression-scheme point of view, as it has $N=M$ and hence achieves no compression. However, it strongly indicates that the conditions \eqref{eq:constraintC} and \eqref{eq:constraintI} are not sufficient to ensure the existence of a set construction with the four required properties. Nonetheless, valid set constructions have been found for small values of $N$, $M$ and $R$; for example, $(N,M,R)=(10,6,3)$ yields $C=5$, $I=2$, and the set construction
\begin{eqnarray}
\mathbf{U}_1&=&\left\{1,2,3,4,5\right\},\nonumber\\
\mathbf{U}_2&=&\left\{1,2,6,7,8\right\},\nonumber\\
\mathbf{U}_3&=&\left\{1,3,6,9,10\right\},\nonumber\\
\mathbf{U}_4&=&\left\{2,5,8,9,10\right\},\nonumber\\
\mathbf{U}_5&=&\left\{3,4,7,8,10\right\},\nonumber\\
\mathbf{U}_6&=&\left\{4,5,6,7,9\right\}.
\end{eqnarray}
Additional solutions for $(N,M,R)=(12,9,3)$ and $(N,M,R)=(14,7,3)$ also exist. No counterexamples (i.e. values of $(N,M,R)$ satisfying \eqref{eq:constraintC} and \eqref{eq:constraintI} but admitting no set construction) have been found for $N>M$, although we have not conducted an exhaustive search.

A general compression scheme satisfying the conditions \eqref{eq:constraintC} and \eqref{eq:constraintI} might potentially admit more compression rates than the symmetric base scheme for each value of $N$. Given the difficulties in actually constructing the sets, however, we focus instead on the special case of ``maximal representation'' for fixed $M$ and $R$ (i.e. every $M$-digit binary number with exactly $R$ 1's represents a distinct template label); this gives the binomial coefficient scheme described in Sec.\,\ref{subsec:binomial}.

\section{Taylor-T2 PN expansions}\label{app:PN}

The Taylor-T2 PN waveform \eqref{eq:taylort2} used in Sec.\,\ref{sec:example} describes the inspiral part of a circular and non-inclined comparable-mass binary merger \cite{BIWW1996,B2014,CG2014}. Its amplitude and phase are written as expansions in the frequency-related variable
\begin{equation}
x(t)=\left(\frac{G\mathcal{M}}{c^3\eta^{3/5}}\frac{d}{dt}\phi(t)\right)^{2/3},
\end{equation}
with the orbital phase $\phi$ given to 3.5PN accuracy by

\begin{widetext}
\begin{multline}
\phi(t)=-\frac{1}{\eta}\Bigg\{\tau^{5/8}
+\left(\frac{3715}{8064}+\frac{55}{96}\eta\right)\tau^{3/8}
-\frac{3}{4}\pi\tau^{1/4}
+\left(\frac{9275495}{14450688}+\frac{284875}{258048}\eta+\frac{1855}{2048}\eta^2\right)\tau^{1/8}\\
+\left(-\frac{38645}{172032}+\frac{65}{2048}\eta\right)\pi\ln{\left(\frac{\tau}{\tau(0)}\right)}
+\left(\frac{831032450749357}{57682522275840}-\frac{53}{40}\pi^2-\frac{107}{56}\gamma+\frac{107}{448}\ln{\left(\frac{\tau}{256}\right)}\right.\\
\left.+\left(-\frac{126510089885}{4161798144}+\frac{2255}{2048}\pi^2\right)\eta+\frac{154565}{1835008}\eta^2-\frac{1179625}{1769472}\eta^3\right)\tau^{-1/8}\\
+\left(\frac{188516689}{173408256}+\frac{488825}{516096}\eta-\frac{141769}{516096}\eta^2\right)\pi\tau^{-1/4}\Bigg\},
\end{multline}
\end{widetext}

\noindent where $\gamma$ is the Euler--Mascheroni constant. Here $\tau$ is a time-related variable, written in terms of the binary coalescence time $t_c$ as
\begin{equation}
\tau(t)=\frac{c^3\eta^{8/5}}{5G\mathcal{M}}(t_c-t),
\end{equation}
where we set $t_c=1\,\mathrm{yr}$ for a massive ($\sim10^6M_\odot$) black-hole binary inspiral.

The GW amplitude is then proportional to the 2PN amplitude function
\begin{multline}
A=x\left(2+\frac{1}{3}(-13+\eta)x+4\pi x^{3/2}\right.\\\left.+\frac{1}{180}(-837-635\eta+15\eta^2)x^2\right),
\end{multline}
while the GW phase is twice the tail-distorted orbital phase
\begin{equation}
\psi=\phi-3x^{3/2}\left(1-\frac{\eta}{2}x\right)\ln{\left(\frac{x}{x(0)}\right)},
\end{equation}
with the 1PN factor of $1-(\eta/2)x$ included to account for the nonlinear interaction between the gravitational field of the source and its emitted gravitational radiation \cite{BS1993}. The constant frequency in $x(0)$ is set to $\dot{\phi}(0)=10^{-4}\pi$, which corresponds to an approximate entry frequency of $10^{-4}\,\mathrm{Hz}$ for the eLISA detector \cite{AEA2013}.

\bibliographystyle{unsrt}
\bibliography{references}
\end{document}